\documentclass[12pt,sort&compress]{elsarticle}

\usepackage{afterpage}
\usepackage{enumitem}
\usepackage{mystyle}
\usepackage{algpseudocode}
\usepackage{algorithm}
\usepackage{amsmath}
\usepackage{amssymb}

\DeclareMathOperator*{\argmin}{arg\,min}

\begin{document}

\begin{frontmatter}

\title{$S$-Frame Discrepancy Correction Models for\\ Data-Informed Reynolds Stress Closure}

\author[ball]{Eric L. Peters}
\author[colorado]{Riccardo Balin}
\author[colorado]{Kenneth E. Jansen}
\author[colorado]{Alireza Doostan}
\author[colorado]{John~A.~Evans\corref{cor1}}
\ead{john.a.evans@colorado.edu}
\cortext[cor1]{Corresponding author}

\address[ball]{Ball Aerospace \& Technologies Boulder, CO 80301, USA}
\address[colorado]{University of Colorado Boulder, Boulder, CO 80309, USA}

\begin{abstract}
Despite their well-known limitations, Reynolds averaged Navier-Stokes (RANS) models remain the most commonly employed tool for modeling turbulent flows in engineering practice.  RANS models are predicated on the solution of the RANS equations, but the RANS equations involve an unclosed term, the Reynolds stress tensor, which must be modeled.  The Reynolds stress tensor is often modeled as an algebraic function of mean flow field variables and turbulence variables.  This, however, introduces a discrepancy between the Reynolds stress tensor predicted by the Reynolds stress model and the exact Reynolds stress tensor.  This discrepancy can result in inaccurate mean flow field predictions for complex flows of industrial relevance.  In this paper, we introduce a data-informed approach for arriving at Reynolds stress models with improved predictive performance.  Our approach relies on learning the components of the Reynolds stress discrepancy tensor associated with a given Reynolds stress model in the mean strain-rate tensor eigenframe.  These components are typically smooth and hence simple to learn using state-of-the-art machine learning strategies and regression techniques.  Our approach automatically yields Reynolds stress models that are symmetric, and it yields Reynolds stress models that are both Galilean and frame invariant provided the inputs are themselves Galilean and frame invariant.  To arrive at computable models of the discrepancy tensor, we employ feed-forward neural networks and an input space spanning the integrity basis of the mean strain-rate tensor, the mean rotation-rate tensor, the mean pressure gradient, and the turbulent kinetic energy gradient, and we introduce a framework for dimensional reduction of the input space to further reduce computational cost.  Numerical results illustrate the effectiveness of the proposed approach for data-informed Reynolds stress closure for a suite of turbulent flow problems of increasing complexity.
\end{abstract}


\end{frontmatter}

%\author{Eric L. Peters, Riccardo Balin, Kenneth E. Jansen, Alireza Doostan and John A. Evans}
%\maketitle

\section{Introduction}
\label{sec:introduction}
Turbulence evolves through highly nonlinear interaction of a broad spectrum of spatial and temporal scales.  Resolving all of these scales with direct numerical simulation (DNS) can require the largest supercomputer for months for a relatively simple flow \cite{moin1998direct}.  Scale-resolving simulation (SRS) methodologies such as large eddy simulation (LES) \cite{lesieur1996new} and detached eddy simulation (DES) \cite{spalart2009detached} are substantially more economical than DNS in terms of computational cost, but they are still too expensive for use in applications such as design optimization and uncertainty quantification where flow predictions for several problem-defining parameter combinations are required.  As such, Reynolds averaged Navier-Stokes (RANS) models remain the industry standard for turbulence modeling \cite{corson2009industrial}.

RANS models are predicated on the solution of the RANS equations which govern the dynamics of the mean flow field.  Unfortunately, the RANS equations involve an unclosed term, the Reynolds stress tensor, which cannot be expressed in terms of the mean velocity and pressure fields.  It is possible to derive governing transport equations for the Reynolds stress tensor, but this gives rise to even more unclosed terms \cite{launder1975progress}.  It is far more common to model the Reynolds stress tensor as an algebraic function of mean flow field variables and turbulence variables (e.g., turbulent kinetic energy and turbulent kinetic energy dissipation rate) and introduce model transport equations for the turbulence variables using physics-based insight.  Linear eddy viscosity (LEV) models \cite{launder1974numerical,wilcox1988reassessment,spalart1992one,menter1994two}, nonlinear eddy viscosity (NLEV) models \cite{pope1975more,speziale1987nonlinear}, and explicit algebraic Reynolds stress (EARS) models \cite{gatski1993explicit} are all constructed in such a fashion.  However, there are two types of modeling error associated with these approaches: i) there are discrepancies between the model and exact closure terms (e.g., turbulent transport and pressure transport) appearing in the model transport equations and ii) there is a discrepancy between the Reynolds stress tensor predicted by the Reynolds stress model and the exact Reynolds stress tensor \cite{xiao2019quantification}.  These modeling errors can result in inaccurate mean flow field predictions for complex flows of industrial relevance \cite{pope2001turbulent}.

The data-informed approach\footnote{The data-informed approach to RANS model closure is often referred as data-driven rather than data-informed.  However, data is typically used to supplement rather than replace \textit{a priori} knowledge in RANS model closure, so we prefer the term data-informed over data-driven.} to RANS model closure has recently arisen as an attractive candidate for overcoming the deficiencies of state-of-the-art RANS models \cite{tracey2013application,tracey2015machine,parish2016paradigm,ling2016reynolds,xiao2016quantifying,wang2017physics,wu2018physics,wu2019representation,zhu2019machine,edeling2018data,schmelzer2020discovery}.  This approach leverages advances in machine learning and the availability of high-fidelity data from simulations and experiments to build improved RANS models.  While it is possible to build a data-informed model from scratch, as in \cite{zhu2019machine}, it is often preferable to start with a particular state-of-the-art RANS model with fully specified model constants.  Model discrepancies are then assessed for the specified RANS model by comparing model closures with exact closures attained using available high-fidelity data \cite{wang2017physics,wu2019representation,schmelzer2020discovery}.  Finally, computable models of the model discrepancies in terms of a chosen set of input features are obtained using supervised machine learning strategies and regression techniques such as neural networks \cite{ling2016reynolds} and random forests \cite{wang2017physics}.

\begin{figure}[t!]
  \centering
  \includegraphics[scale=1.0]{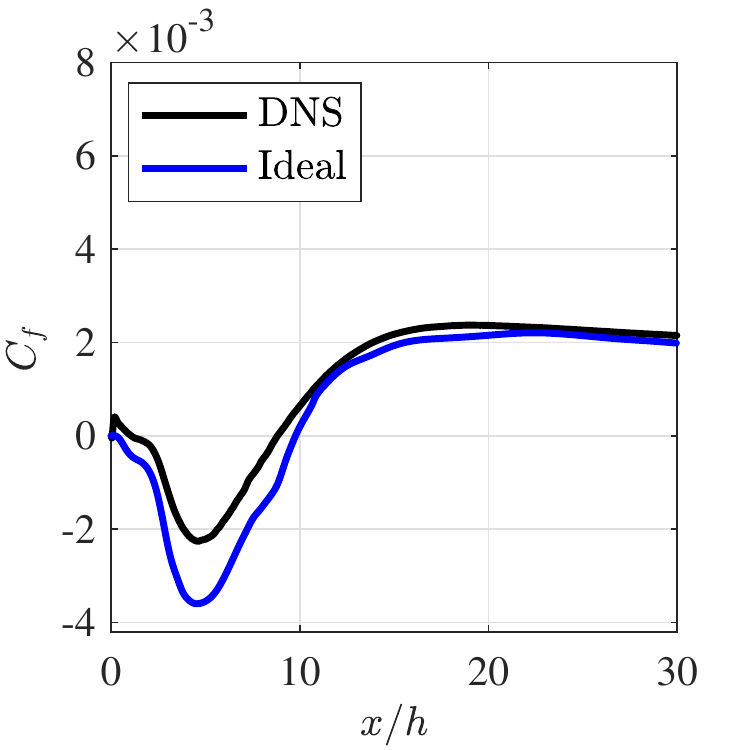}
\caption{Backward-facing step flow at $Re_{\tau} = 395$ and expansion ratio 2: The skin friction coefficient along the bottom wall as predicted using DNS and the ideal viscosity model.  DNS data acquired from \cite{pont2019direct}.}
\label{Figure1}
\end{figure}

There is a substantial and growing literature on the data-informed approach to RANS model closure, and a number of works have focused on learning the discrepancies between the model and exact closure terms appearing in model transport equations.  For example, neural networks were employed in \cite{tracey2015machine} to model the discrepancy in the source term appearing in the Spalart-Allmaras turbulence model, and Gaussian processes were employed in \cite{parish2016paradigm} to model the discrepancy in the production term in the $k$-$\omega$ turbulence model.  However, it is arguably of equal or greater importance to learn the discrepancy between the Reynolds stress tensor predicted by a given Reynolds stress model and the exact Reynolds stress tensor.  This is especially true for LEV models, as there exists no universally accurate LEV model \cite{coleman2018numerical}.

To demonstrate that there exists no LEV model that is universally accurate, consider turbulent flow with inflow friction Reynolds number $Re_{\tau} = 395$ over a backward-facing step with expansion ratio 2.  This is a notoriously challenging problem to model due to the strong flow separation which results from the stress singularity that occurs at the re-entrant corner, and DNS data for this problem was only just recently reported in \cite{pont2019direct}.  We have computed an ideal eddy viscosity from the exact Reynolds stress tensor (attained using the DNS data) in a point-wise manner via the relation
\begin{equation}
\nu_{t_{\text{ideal}}} = \argmin_{\nu_t \in \mathbb{R}_{\geq 0}} \sum_{i,j} \frac{1}{2} \left( a_{ij} + 2\nu_t {S}_{ij} \right)^2, \nonumber
\end{equation}
where $a_{ij}$ is the anistropic part of the Reynolds stress tensor and ${S}_{ij}$ is the mean strain-rate tensor, and we have plotted the skin friction coefficient along the bottom wall as predicted with an LEV model using this ideal eddy viscosity alongside the exact skin friction coefficient in Figure \ref{Figure1}.  Note there is a significant difference between the predicted and exact skin friction coefficients, and in particular, the ideal eddy viscosity model predicts reattachment at a much later location than the DNS reattachment point.  Since the ideal eddy viscosity model best approximates the exact Reynolds stress tensor among all possible LEV models, this demonstrates there exists no LEV model which can accurately predict the mean flow field over the backward facing step.  This further suggests there is limited value in using machine learning strategies to arrive at improved LEV models.

Several different approaches for learning the discrepancy between the Reynolds stress tensor predicted by a given Reynolds stress model and the exact Reynolds stress tensor have been explored in the literature, each with their own set of advantages and disadvantages.  The most obvious approach is to learn the individual components of the discrepancy tensor with respect to a global coordinate system.  The downside of such an approach, however, is that it results in a Reynolds stress model that is not frame invariant.  An alternative approach is to learn the eigenstructure (i.e., the eigenvalues and eigenvectors) of the discrepancy tensor \cite{wang2017physics}.  One pronounced advantage of this approach is that realizability can be enforced by restricting the eigenvalues to lie within the Lumley triangle \cite{lumley1979computational}. Additionally, if the discrepancy tensor eigenvectors are represented in terms of a rotation of the mean strain-rate tensor eigenframe, one arrives at a model that is frame invariant provided the inputs to the model are invariant as well.  Such a rotation can be represented using Euler angles, but these angles are typically nonsmooth and thus quite difficult to learn using state-of-the-art supervised machine learning strategies and regression techniques.  One can somewhat alleviate this issue using quaternions rather than Euler angles \cite{wu2019representation}, but as we demonstrate later, quaternions are also nonsmooth for certain flow configurations.  A final approach that has been explored is to express the discrepancy tensor in terms of an infinite polynomial expansion of a prescribed set of input tensors \cite{ling2016reynolds}.  Provided the set of input tensors is finite in size, one can employ the Cayley-Hamilton Theorem to convert this infinite polynomial expansion to a finite polynomial expansion in terms of the so-called tensor integrity basis.  When the number of input tensors is small, the size of the tensor integrity basis is reasonable.  Unfortunately, the size of the tensor integrity basis grows exponentially with the number of input tensors.

In this paper, we present a new approach for learning the discrepancy between the Reynolds stress tensor predicted by a given Reynolds stress model and the exact Reynolds stress tensor.  We focus on learning the Reynolds stress discrepancy associated with LEV models, though our approach extends without modification to other classes of Reynolds stress models including NLEV and EARS models.  Our approach is based on learning the individual components of the discrepancy tensor with respect to a field-specific coordinate system, namely the mean strain-rate tensor eigenframe.  As we demonstrate later in this paper, these components are smooth and hence relatively simple to learn.  Moreover, our approach yields Reynolds stress models that are frame invariant provided the inputs are themselves invariant, and it also yields models whose stability characteristics are easily established.  In particular, our approach enables us to construct Reynolds stress models which result in a global decay of mean kinetic energy.  One deficiency of our approach is that it does not automatically yield Reynolds stress models that are realizable.  However, we believe this deficiency to be of minor concern, and we demonstrate that realizability can be directly enforced using a post-processing procedure if desired.  To arrive at computable models of the discrepancy tensor, we turn to feed-forward neural networks and employ an input feature space comprising the integrity basis of the mean strain-rate tensor, the mean rotation-rate tensor, the mean pressure gradient, and the turbulent kinetic energy gradient. We also investigate the use of Sobol indices to perform dimensional reduction of the input feature space while maintaining predictive performance.  

An outline of this paper is as follows.  In the next section, we recall the incompressible Reynolds averaged equations and the Reynolds stress tensor.  In Section 3, we present the different representations of the Reynolds stress tensor and its discrepancy tensor.  Next we go on to describe how we model the discrepancy tensor with our specific representation in Section 4.  In Sections 5 and 6, we discuss how to enforce energetic stability and realizability using our discrepancy model.  We then present numerical results using both specific and universal neural networks to model our representation of the discrepancy tensor in Section 7.  In Section 8, we introduce methods for dimensional reduction in the context of turbulence modeling.  Lastly, in Section 9, we draw conclusions and discuss future research directions.

\section{The RANS equations and the Reynolds Stress Tensor}
\label{sec:rans}
We begin by recalling the RANS equations for an incompressible flow.  The derivation of the RANS equations relies on a \textit{Reynolds decomposition} of the \textit{velocity} and \textit{pressure} fields ${u}_i$ and $p$ into \textit{mean} and \textit{fluctuating components}, viz.,
\begin{equation}
\begin{aligned}
{u}_i &= \overline{{u}}_i + {u}_i',\\
p &=  \overline{p}  + p',
\end{aligned} 
\label{mean}
\end{equation}
where $\overline{*}$ and $*'$ denote mean and fluctuation, respectively.  In particular, the RANS equations are obtained by taking the mean of the Navier-Stokes equations and exploiting the above Reynolds decompositions.  The resulting system is displayed below:
\begin{equation}
\begin{aligned}
 \overline{u}_{i,t} +  \overline{{u}}_j \overline{u}_{i,j} + &\frac{1}{\rho}  \overline{p}_{,i} -   \left( 2 \nu S_{ij} \right)_{,j} +  \tau_{ij,j} = \overline{f}_i, \\
&\overline{{u}}_{i,i} = 0.
\end{aligned} 
\label{NS}
\end{equation}
Above, $\rho$ denotes the \textit{density}, $\nu$ denotes the \textit{kinematic viscosity}, $\overline{f}_i$ denotes the \textit{mean body force}, $S_{ij} = \frac{1}{2}(\overline{u}_{i,j} + \overline{u}_{j,i})$ denotes the \textit{mean strain-rate tensor} and $ \tau_{ij} = \overline{{u}_{i}' {u}_{j}'} = \overline{ {u}_i {u}_j } - \overline{u}_i \overline{u}_j$ denotes the \textit{Reynolds stress tensor}.  The Reynolds stress tensor cannot be expressed directly in terms of the mean velocity and pressure fields, so typically it is modeled with a \textit{Reynolds stress closure}.  The Reynolds stress tensor admits several important properties which should ideally be preserved by a given Reynolds stress closure.  In particular, the Reynolds stress tensor is (1) \textit{symmetric}, (2) \textit{realizable} (that is, its eigenvalues are non-negative), (3) \textit{frame invariant} (that is, it maps as a tensor under a change of coordinates), and (4) \textit{Galilean invariant} (that is, it is invariant under Galilean transformations).

\section{Reynolds Stress Discrepancies in Linear Eddy Viscosity Models}
\label{sec:discrepancy}
\begin{figure}[t!]
	\begin{subfigure}[b]{1.0\linewidth}
  		\includegraphics[width=1.0\linewidth]{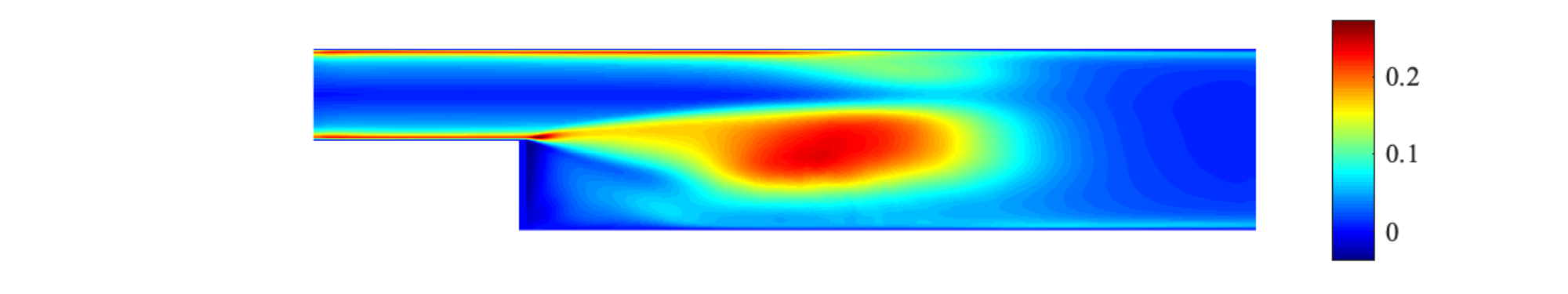}
  		\caption{$a_{11}$ ($m^2/s^2$)}
	\end{subfigure}
	\begin{subfigure}[b]{1.0\linewidth}
  		\includegraphics[width=1.0\linewidth]{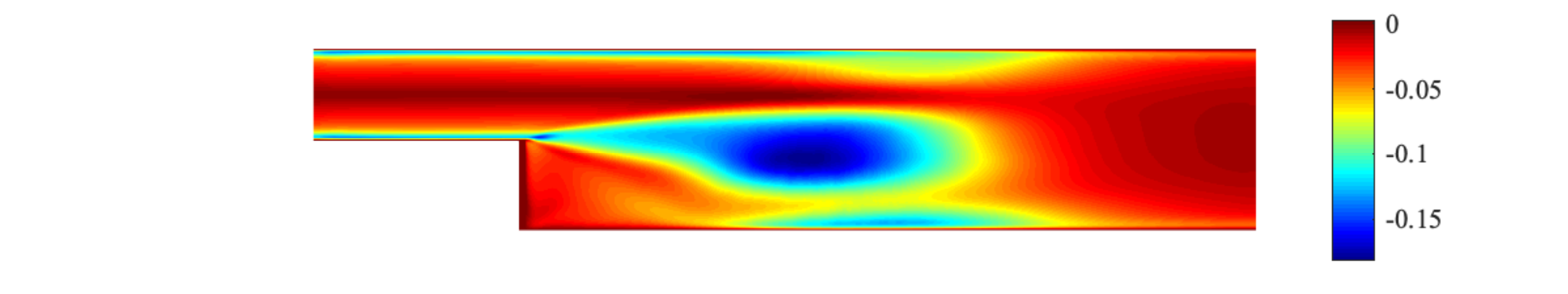}
  		\caption{$a_{22}$ ($m^2/s^2$)}
	\end{subfigure}
	\begin{subfigure}[b]{1.0\linewidth}
  		\includegraphics[width=1.0\linewidth]{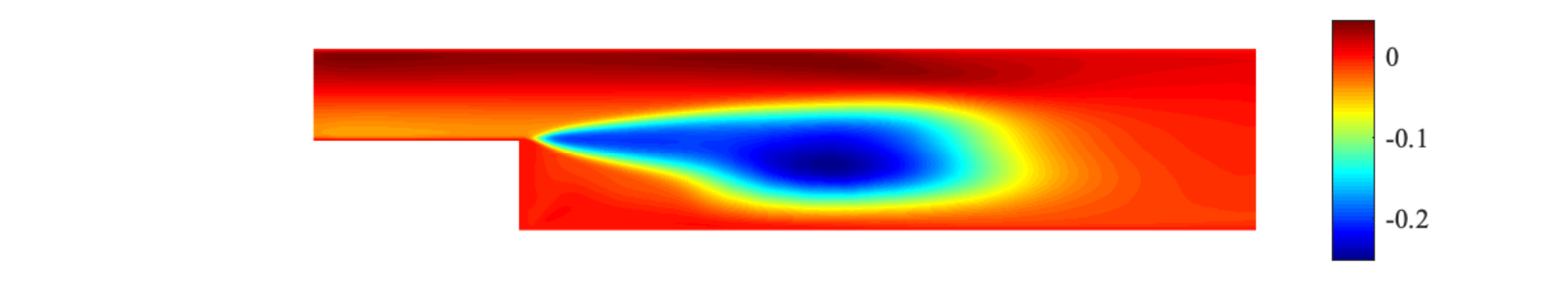}
  		\caption{$a_{12}$ ($m^2/s^2$)}
	\end{subfigure}
	\caption{Components of the anisotropic Reynolds stress tensor $a_{ij}$ in the global coordinate frame.}
	\label{Figure2}
\end{figure}

\begin{figure}[t!]
\centering
    \begin{subfigure}[b]{1.0\linewidth}
  		\includegraphics[width=1.0\linewidth]{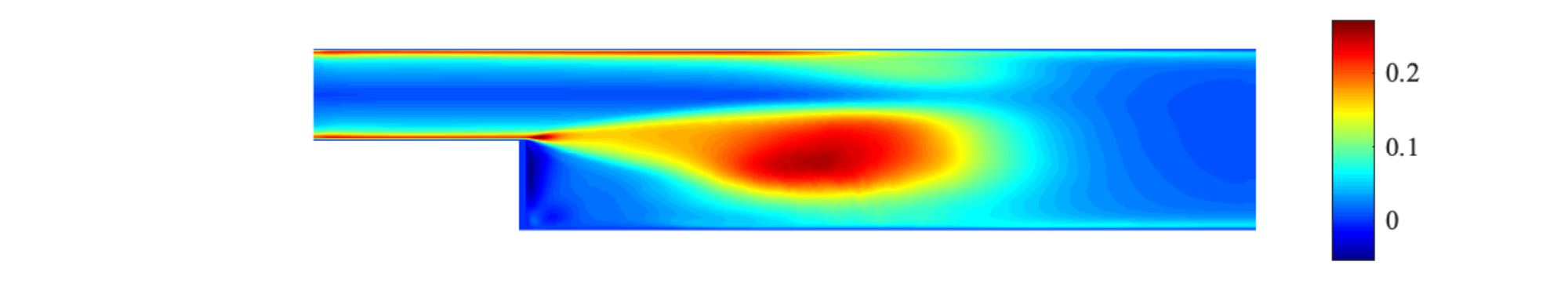}
  		\caption{$D_{11}$ ($m^2/s^2$)}
	\end{subfigure}
	\begin{subfigure}[b]{1.0\linewidth}
  		\includegraphics[width=1.0\linewidth]{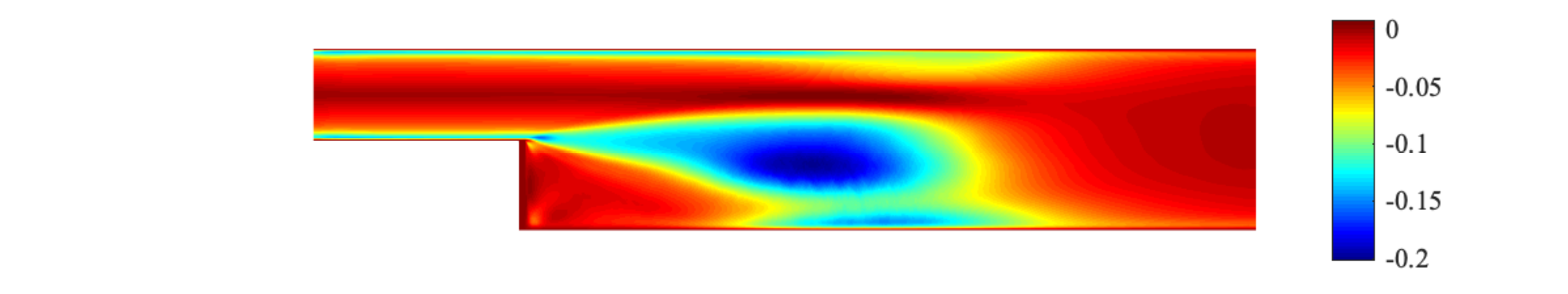}
  		\caption{$D_{22}$ ($m^2/s^2$)}
	\end{subfigure}
	\begin{subfigure}[b]{1.0\linewidth}
  		\includegraphics[width=1.0\linewidth]{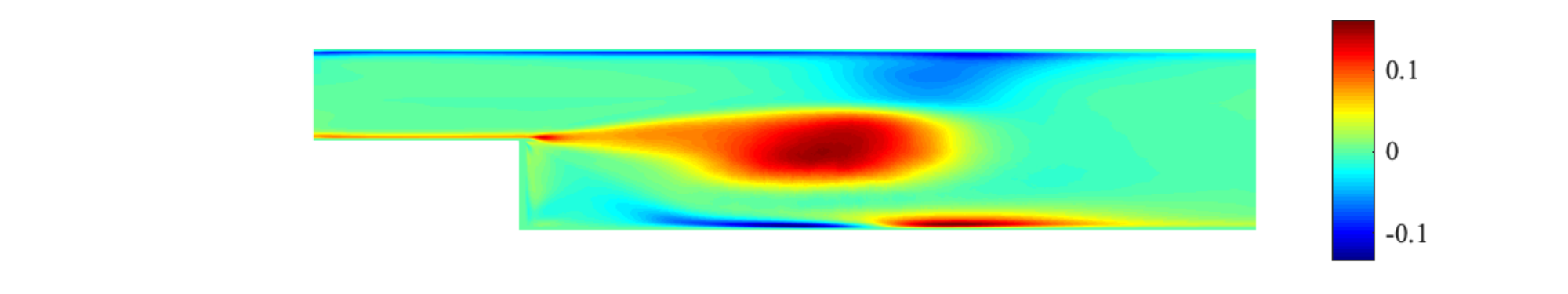}
  		\caption{$D_{12}$ ($m^2/s^2$)}
	\end{subfigure}
	\caption{Components of the discrepancy tensor $D_{ij}$ associated with the standard $k$-$\varepsilon$ model in the global coordinate frame.}
	\label{Figure3}
\end{figure}

The most common approach to Reynolds stress closure is to employ a \textit{linear eddy viscosity} (LEV) model \cite{launder1974numerical,wilcox1988reassessment,spalart1992one,menter1994two}.  This approach begins by decomposing the Reynolds stress tensor as
\begin{equation}
\tau_{ij} = a_{ij} + \frac{2}{3} k \delta_{ij},
\end{equation}
where $a_{ij}$ denotes the \textit{anisotropic} part of the tensor whose trace is equal to zero and $\frac{2}{3} k \delta_{ij}$ denotes the \textit{isotropic} part of the tensor whose trace is equal to twice the \textit{turbulent kinetic energy} $k = \frac{1}{2} \overline{{u}_{i}' {u}_{i}'}$.  The anisotropic part of the Reynolds stress tensor is then modeled according to
\begin{equation}
a_{ij} \approx -2 \nu_t S_{ij},
\end{equation}
where $\nu_t$ is a \textit{turbulence eddy viscosity}.  The turbulence eddy viscosity is typically related to the mean flow field and one or more turbulence variables.  For instance, the turbulence eddy viscosity associated with the standard $k$-$\varepsilon$ model is
\begin{equation}
\nu_t = C_{\mu} \frac{k^2}{\varepsilon},
\end{equation}
where $\varepsilon = 2 \nu \langle s_{ij} s_{ij} \rangle$ is the \textit{turbulent kinetic energy dissipation rate}, $s_{ij} = \frac{1}{2} \left( u'_{i,j} + u'_{j,i} \right)$ is the \textit{fluctuating strain-rate tensor}, and $C_{\mu}$ is a model constant that is typically set to 0.09 \cite{launder1974numerical}.  Typically, LEV models are imperfect and thus there is a discrepancy between the Reynolds stress tensor predicted by the model and the exact Reynolds stress tensor.  We represent this discrepancy using a \textit{discrepancy tensor},
\begin{equation}
D_{ij} = a_{ij} + 2 \nu_t S_{ij}.
\end{equation}
The Reynolds stress tensor then admits the \textit{exact} decomposition,
\begin{equation}
\tau_{ij} = -2 \nu_t S_{ij} + D_{ij} + \frac{2}{3} k \delta_{ij}.
\end{equation}
To visualize the above concepts, we turn again to the turbulent flow problem that was considered in Section \ref{sec:introduction}.  Namely, we consider turbulent flow with inflow friction Reynolds number $Re_{\tau} = 395$ over a backward-facing step with expansion ratio 2.  In Figure \ref{Figure2}, the three unique components of the anisotropic tensor are displayed with respect to a global coordinate frame whose $x$-axis is aligned with the bottom wall and $y$-axis is aligned with the left wall.  In Figure \ref{Figure3}, the corresponding components of the discrepancy tensor associated with the standard $k$-$\varepsilon$ model are displayed.  Note that the components of the discrepancy tensor are nonzero near the walls as well as in the separating region of the flow.  While the predictive performance of the standard $k$-$\varepsilon$ model near the walls may be improved using a near-wall treatment such as wall damping \cite{chien1982predictions}, its performance in the separating region of the flow cannot be improved by simply modifying the turbulence eddy viscosity.  This is due to the fact that even an ideal LEV model fails to accurately model this flow, as discussed in Section \ref{sec:introduction}.  Consequently, to improve the performance of the standard $k$-$\varepsilon$ model throughout the flow, one must construct a suitable model for the discrepancy tensor.

\begin{figure}[t!]
    \begin{subfigure}[b]{1.0\linewidth}
  		\includegraphics[width=1.0\linewidth]{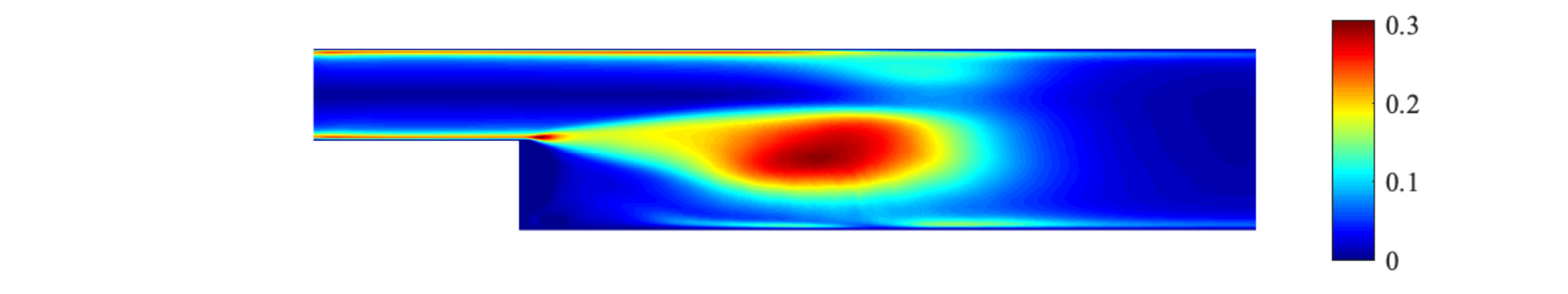}
  		\caption{First in-plane eigenvalue ($m^2/s^2$)}
	\end{subfigure}
	\begin{subfigure}[b]{1.0\linewidth}
  		\includegraphics[width=1.0\linewidth]{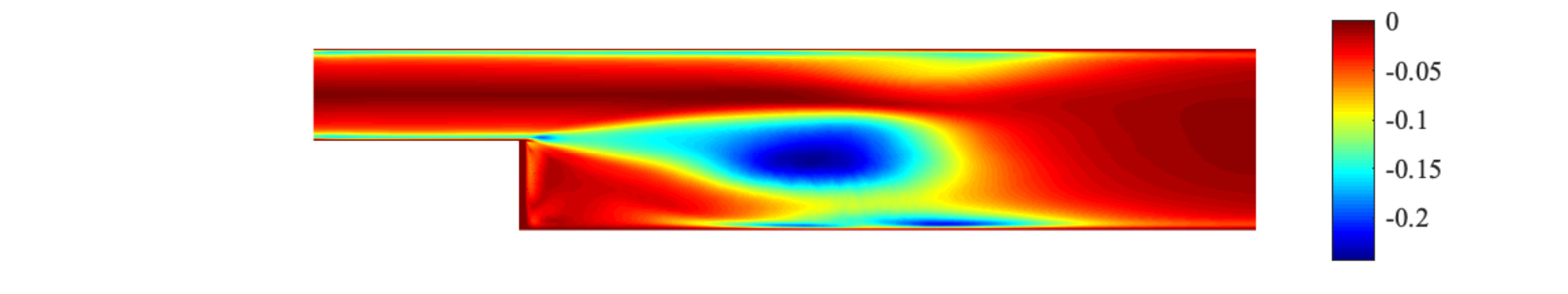}
  		\caption{Second in-plane eigenvalue ($m^2/s^2$)}
	\end{subfigure}
	\begin{subfigure}[b]{1.0\linewidth}
  		\includegraphics[width=1.0\linewidth]{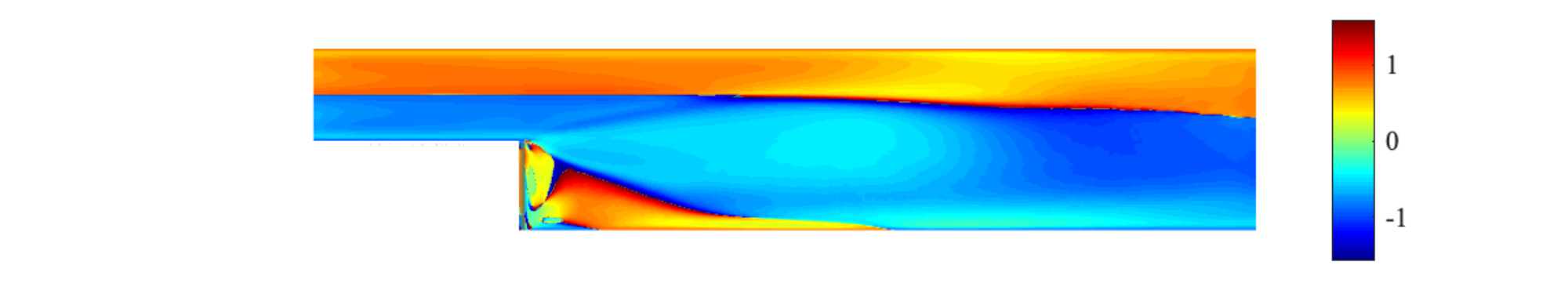}
  		\caption{Euler angle about z-axis ($rad$)}
	\end{subfigure}
	\caption{Eigenvalues of the discrepancy tensor $D_{ij}$ associated with the standard $k$-$\varepsilon$ model and Euler angles of the rotation matrix $R^{S \rightarrow D}_{ij}$.}
	\label{Figure4}
\end{figure}

The most straightforward means of modeling the discrepancy tensor is to model each of its components with respect to the global coordinate frame.  Unfortunately, such an approach will result in a model that is \textit{not} frame invariant.  This is due to the fact that the individual components of the discrepancy tensor change under a change of coordinates.  That is, the individual components of the discrepancy tensor are not scalars.

\begin{figure}[t!]
    \begin{subfigure}[b]{1.0\linewidth}
  		\includegraphics[width=1.0\linewidth]{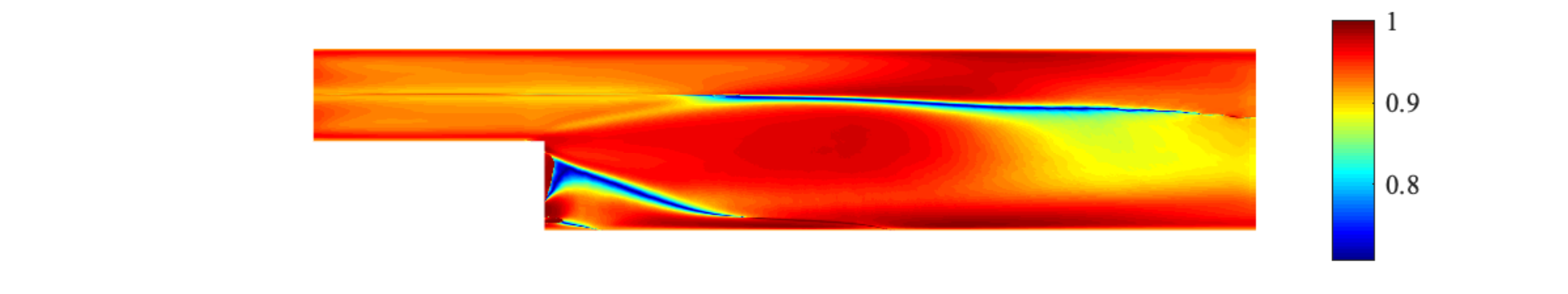}
  		\caption{First quaternion component}
	\end{subfigure}
	\begin{subfigure}[b]{1.0\linewidth}
  		\includegraphics[width=1.0\linewidth]{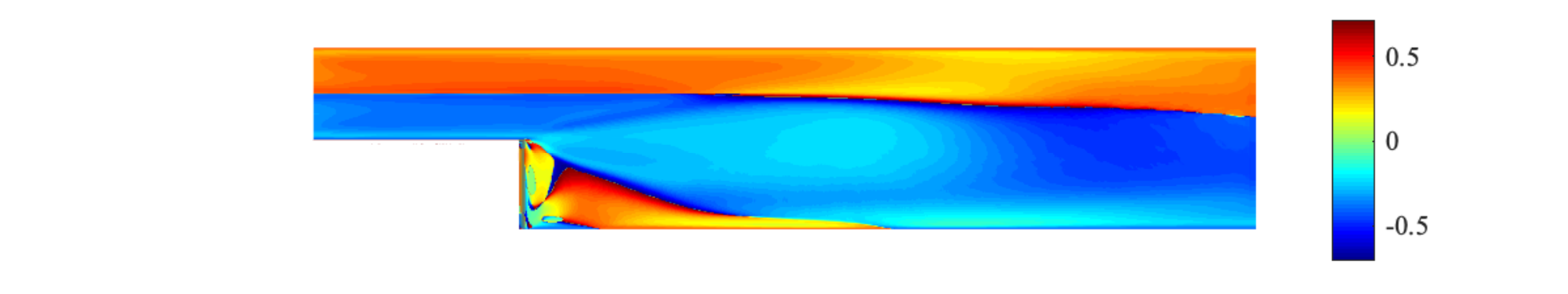}
  		\caption{Second quaternion component}
	\end{subfigure}
	\caption{Nonzero quaternion components associated with the rotation matrix $R^{S \rightarrow D}_{ij}$.}
	\label{Figure5}
\end{figure}

An alternative means of modeling the discrepancy tensor is to model its \textit{eigendecomposition},
\begin{equation}
D_{ij} = V^D_{ik} \Lambda^D_{kl} V^D_{jl},
\end{equation}
where $V^D_{ij}$ is a matrix whose columns are orthonormal \textit{eigenvectors} of $D_{ij}$ satisfying a right-hand-rule (expressed in terms of the global coordinate system) and $\Lambda^D_{ij}$ is a diagonal matrix whose diagonal entries are the \textit{eigenvalues} of $D_{ij}$.  This approach was proposed in \cite{wang2017physics}.  As opposed to the individual components of the discrepancy tensor, the eigenvalues of the discrepancy tensor do not change under a change of coordinates.  Moreover, an eigendecomposition-based discrepancy model can be made traceless by enforcing the sum of the eigenvalues to be equal to zero, and it can be made realizable by enforcing the eigenvalues to lie within the \textit{Lumley triangle} \cite{lumley1979computational}.  However, the main obstacle associated with an eigendecomposition-based discrepancy model is representing the orthonormal eigenvectors of the discrepancy tensor.  One naive approach is to represent the individual components of the eigenvectors.  However, this approach requires learning a model for nine different quantities.  Moreover, it does not result in a frame invariant model as the individual components of the eigenvectors change under a change of coordinates.  An alternative approach is based on the recognition that the matrix $V^D_{ij}$ is an \textit{orthogonal matrix} with \textit{determinant one}. Therefore, it can be represented in terms of three Euler angles.  However, these angles are again with respect to the global coordinate system, so this approach also does not result in a frame invariant model.  Yet another alternative approach is based on the insight that one can relate the eigendecomposition of the discrepancy tensor to the eigendecomposition of the mean strain-rate tensor,
\begin{equation}
S_{ij} = V^S_{ik} \Lambda^S_{kl} V^S_{jl}.
\end{equation}
In particular, one can represent the orthonormal eigenvectors of $D_{ij}$ by rotating the orthonormal eigenvectors of $S_{ij}$ using a \textit{rotation matrix} $R^{S \rightarrow D}_{ij}$,
\begin{equation}
V^D_{ij} = R^{S \rightarrow D}_{ik} V^S_{kj}.
\end{equation}
The rotation matrix $R^{S \rightarrow D}_{ij}$ can also be expressed in terms of three Euler angles.  However, as the rotation matrix $R^{S \rightarrow D}_{ij}$ is with respect to the coordinate frame described by the eigenvectors of $S_{ij}$ rather than the global coordinate system, its Euler angles are frame invariant.  Thus, an eigendecomposition-based discrepancy model based on the eigenvalues of $D_{ij}$ and the Euler angles of $R^{S \rightarrow D}_{ij}$ has the potential of being frame invariant.  However, such a model does suffer from one striking drawback, which is best illustrated by example.  In Figure \ref{Figure4}, the eigenvalues of $D_{ij}$ and the Euler angles of $R^{S \rightarrow D}_{ij}$ are displayed for the backward-facing step flow problem.  Note that we have only plotted the two eigenvalues associated with the in-plane eigenvectors as the third may be found from these two, and we have only plotted one Euler angle as the other two are zero for this statistically two-dimensional flow.  While the two in-plane eigenvalues of $D_{ij}$ are smooth throughout the flow domain, the displayed Euler angle is remarkably non-smooth, especially in the separating region of the flow.  This nonsmoothness renders the Euler angle very difficult to learn using state-of-the-art machine learning approaches.  It was recently proposed in \cite{wu2019representation} to represent the rotation matrix $R^{S \rightarrow D}_{ij}$ using a quaternion as quarterions typically exhibit enhanced smoothness as compared with Euler angles.  Moreover, quaternions do not suffer from gimbal locking.  However, the first two components of the quaternion associated with $R^{S \rightarrow D}_{ij}$ are displayed in Figure \ref{Figure5} for the backward-facing step flow problem, and it is clearly evident that the two components are not smooth.

\begin{figure}[t!]
\centering
	\begin{subfigure}[b]{1.0\linewidth}
  		\includegraphics[width=1.0\linewidth]{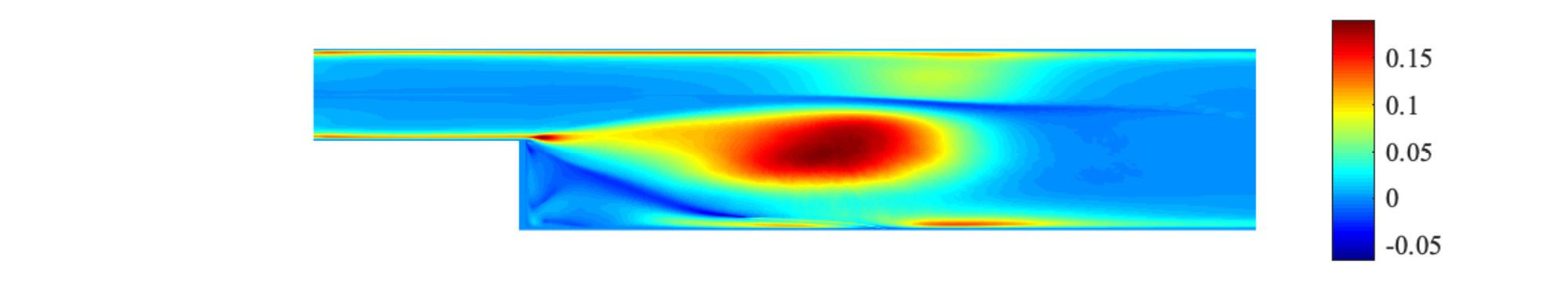}
  		\caption{$D^S_{11}$ ($m^2/s^2$)}
	\end{subfigure}
	\begin{subfigure}[b]{1.0\linewidth}
  		\includegraphics[width=1.0\linewidth]{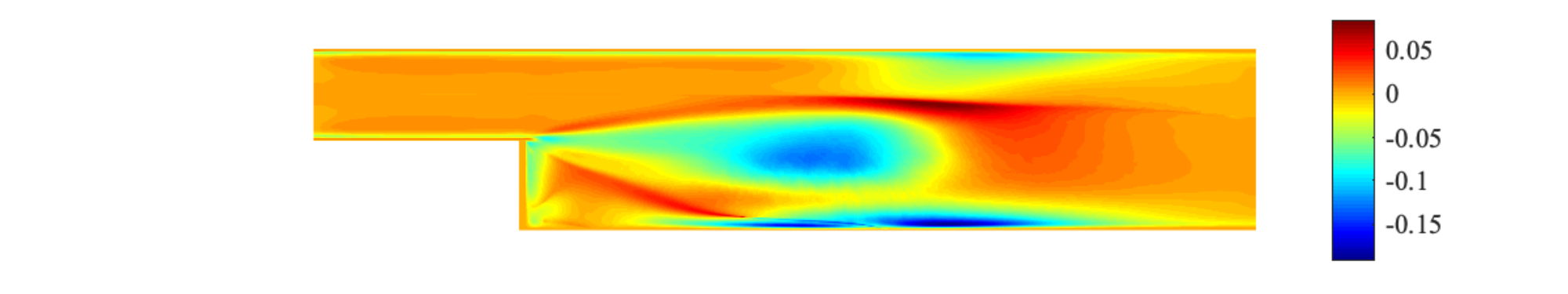}
  		\caption{$D^S_{22}$ ($m^2/s^2$)}
	\end{subfigure}
	\begin{subfigure}[b]{1.0\linewidth}
  		\includegraphics[width=1.0\linewidth]{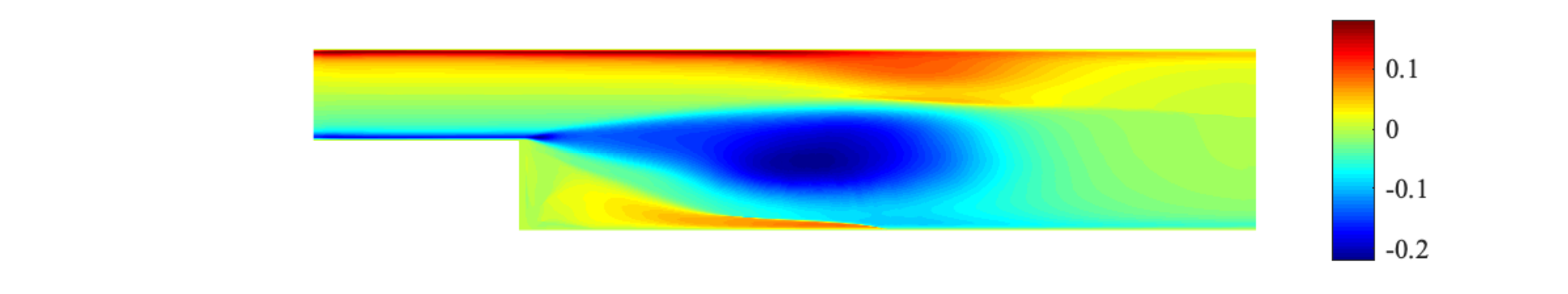}
  		\caption{$D^S_{12}$ ($m^2/s^2$)}
	\end{subfigure}
	\caption{Components of the discrepancy tensor $D_{ij}$ associated with the standard $k$-$\varepsilon$ model in the coordinate frame associated with the eigenvectors of the mean strain-rate tensor $S_{ij}$.}
	\label{Figure6}
\end{figure}

Given the difficulties associated with eigendecomposition-based discrepancy models, we propose here a new approach for modeling the discrepancy tensor.  Our approach is based on representing the components of the discrepancy tensor in the coordinate frame associated with the eigenvectors of the mean strain-rate tensor.  We refer to these components as $D^S_{ij}$.  The components of the discrepancy tensor in the global coordinate frame may be recovered via
\begin{equation}
D_{ij} = V^S_{ik} D^S_{kl} V^S_{jl}.
\end{equation}
Consequently, the full Reynolds stress tensor admits the exact decomposition,
\begin{equation}
\tau_{ij} = -2 \nu_t S_{ij} + V^S_{ik} D^S_{kl} V^S_{jl} + \frac{2}{3} k \delta_{ij}.
\end{equation}
We refer to discrepancy models based on modeling the components of $D^S_{ij}$ as \textit{S-frame discrepancy models}.  As the components of $D^S_{ij}$ do not change under a change of coordinates, $S$-frame discrepancy models are frame invariant provided their inputs are also frame invariant.  Moreover, as the discrepancy tensor is symmetric and traceless, one only needs to model five unique components of $D^S_{ij}$.  In Figure \ref{Figure6}, the three in-plane components of $D^S_{ij}$ are displayed.  Note that, as opposed to the Euler angles of the discrepancy tensor, the components of $D^S_{ij}$ are smooth.  Thus, there is potential to learn models of these components using state-of-the-art machine learning techniques.  The one disadvantage of $S$-frame discrepancy models as compared with eigendecomposition-based discrepancy models is that it is quite difficult to construct an \textit{a priori} realizable $S$-frame discrepancy model.  However, we believe that the advantages of $S$-frame discrepancy models in data-driven Reynolds closure overcome this disadvantage, and we illustrate later that realizability can be enforced \textit{a posteriori} using a post-processing procedure if desired.

\begin{figure}[t!]
\centering
	\includegraphics[width=0.45\textwidth]{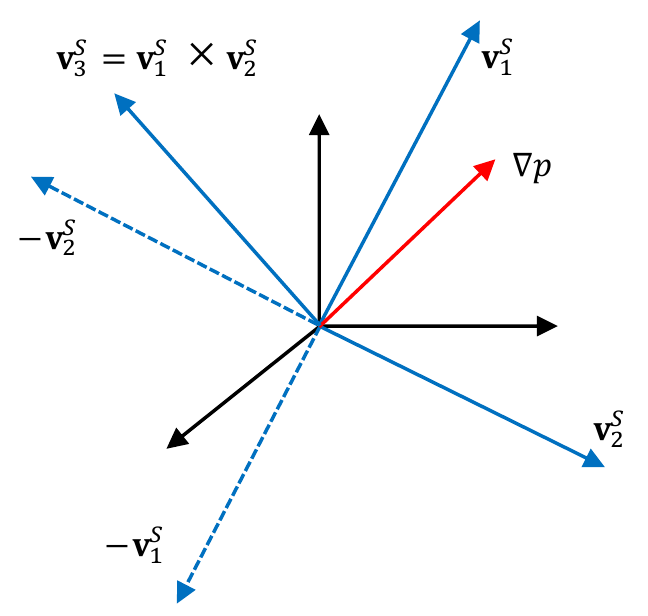}
\caption{Selection of the eigenvectors of the mean strain-rate tensor $S_{ij}$.}
\label{Figure7}
\end{figure}

The vigilant reader may notice that each of the eigenvectors of $S_{ij}$ is unique only up to a sign, and indeed, if any of these signs are flipped, then the values of $D^S_{ij}$ are modified accordingly.  Moreover, if the ordering of the eigenvectors is changed, then the values of $D^S_{ij}$ also change.  As such, a well-posed $S$-frame discrepancy model requires a unique specification of eigenvectors.  We select the first eigenvector of $S_{ij}$ to be that which is most closely aligned with the mean pressure gradient, and we choose the second eigenvector of $S_{ij}$ to be that which is next most closely aligned with the mean pressure gradient.  The third eigenvector of $S_{ij}$ is then obtained by the right-hand-rule.  This is visualized in Figure \ref{Figure7} where the first, second, and third eigenvectors of $S_{ij}$ are denoted as $\textbf{v}^S_1$, $\textbf{v}^S_2$, and $\textbf{v}^S_3$, respectively.

\section{Data-Informed Modeling of the Discrepancy Tensor}
\label{sec:datainformed}
We have so far discussed how to best represent the discrepancy between the Reynolds stress tensor predicted by an LEV model and the exact Reynolds stress tensor.  We have not yet, however, determined how to model this discrepancy.  In this section, we discuss how to construct a symmetric, Galilean invariant, frame invariant, and scale invariant discrepancy model using high-fidelity simulation data.

\subsection{Dimensional Model Form}
\label{subsec:dimensional}
To begin the process of constructing a model for the discrepancy tensor, we must assume a particular dimensional model form.  We assume for the remainder of the paper that the components $D^{S}_{ij}$ of the discrepancy tensor in the $S$-frame may be expressed in terms of the mean strain-rate tensor $S_{ij} = \frac{1}{2}(\overline{u}_{i,j} + \overline{u}_{j,i})$, the mean rotation-rate tensor $\Omega_{ij} = \frac{1}{2}(\overline{u}_{i,j} - \overline{u}_{j,i})$, the mean pressure gradient $\overline{p}_{,i}$, the gradient of turbulent kinetic energy $k_{,i}$, the turbulent kinetic energy $k$, the turbulence time scale $T_t = \frac{k}{\varepsilon}$, and the turbulence eddy viscosity $\nu_t$.  This results in a dimensional $S$-frame discrepancy model of the form
\begin{equation}
D^S_{ij} = D^{S,\text{model}}_{ij}\left( \textbf{S}, \mathbf{\Omega}, \nabla \overline{p}, \nabla k, k, T_t, \nu_t \right),
\label{eq:dimensional_model_form}
\end{equation}
where $\textbf{S} = S_{ij} \textbf{e}_i \otimes \textbf{e}_j$, $\mathbf{\Omega} = \Omega_{ij} \textbf{e}_i \otimes \textbf{e}_j$, $\nabla \overline{p} = \overline{p}_{,i} \textbf{e}_i$, $\nabla k = k_{,i} \textbf{e}_i$, and $\textbf{e}_i$ is the $i^\text{th}$ unit vector.  We refer to $\textbf{S}$, $\mathbf{\Omega}$, $\nabla \overline{p}$, $\nabla k$, $k$, $T_t$, and $\nu_t$ as the \textit{inputs} of the $S$-frame discrepancy model and $D^{S}_{ij}$ as the \textit{outputs} of the $S$-frame discrepancy model.  Note that an $S$-frame discrepancy model of the above form is \textit{Galilean invariant} by construction as the inputs are Galilean invariant.  Finally, it should be mentioned that the procedure discussed in this section may be used to construct $S$-frame discrepancy models with other Galilean invariant inputs (e.g., distance to the wall $d$).  However, such dependencies are not discussed further for ease of exposition.

\subsection{Non-Dimensional Model Form}
\label{subsec:nondimensional}
To arrive at a \textit{scale invariant} $S$-frame discrepancy model, that is, a model independent of unit system, we nondimensionalize both the inputs and outputs.  The Buckingham $\Pi$ theorem may be used for this purpose \cite{bertrand1878homogeneite}.  The advantage of employing the Buckingham $\Pi$ theorem for nondimensionalization is that it also allows one to reduce the dimension of the input space.  The disadvantage of the Buckingham $\Pi$ theorem is that it may yield nondimensional inputs and outputs which vary wildly in size.  This is dentrimental for state-of-the-art machine learning approaches such as deep learning \cite{goodfellow2016deep,higham2019deep}.  We alternatively employ the nondimensionalization procedure proposed in \cite{ling2015evaluation}.  In particular, we nondimensionalize each output $D^S_{ij}$ with the normalization factor $2k$ through the relationship
\begin{equation}
\hat{D}^S_{ij} = \frac{D^S_{ij}}{2k}
\end{equation}
to obtain a corresponding nondimensional output $\hat{D}^S_{ij}$, and we nondimensionalize each input $\alpha$ with a normalization factor $\beta$ through the relationship
\begin{equation}
\hat{\alpha} = \frac{\alpha}{| \alpha | + | \beta |}
\end{equation}
to obtain a corresponding nondimensional input $\hat{\alpha}$.  Following \cite{wang2017physics}, we employ the normalization factors listed in Table \ref{table:nondimensional_inputs} to nondimensionalize the inputs, though alternative normalization factors may also be employed.  The above selections yield a nondimensional $S$-frame discrepancy model of the form
\begin{equation}
\hat{D}^S_{ij} = \hat{D}^{S,\text{model}}_{ij}\left( \hat{\textbf{S}}, \hat{\mathbf{\Omega}}, \widehat{\nabla \overline{p}}, \widehat{\nabla k}, \hat{k}, \hat{T}_t, \hat{\nu}_t \right).
\label{eq:nondimensional_model_form}
\end{equation}
A corresponding dimensional $S$-frame discrepancy model is then obtained via
\begin{equation}
D^{S,\text{model}}_{ij}\left( \textbf{S}, \mathbf{\Omega}, \nabla \overline{p}, \nabla k, k, T_t, \nu_t \right) = 2k \hat{D}^{S,\text{model}}_{ij}\left( \hat{\textbf{S}}, \hat{\mathbf{\Omega}}, \widehat{\nabla \overline{p}}, \widehat{\nabla k}, \hat{k}, \hat{T}_t, \hat{\nu}_t \right).
\end{equation}
Note that, by construction, the magnitudes of the nondimensional inputs are bounded above by one.  We further use a min-max scaling during model training to arrive at nondimensional outputs whose magnitudes are also bounded above by one.  This accelerates the training process and improves the accuracy of the resulting model.

\begin{table}[t!]
\caption{Nondimensionalization of the inputs.}
\centering % used for centering table
\begin{tabular}{|c | c | c|} % centered columns (4 columns)
\hline && \\[-1em]
 Input $\alpha$ & Normalization Factor $\beta$ & Normalized Input $\hat{\alpha}$ \\ [0.5ex] % inserts table
%heading
\hline && \\[-1em]
$\textbf{S}$ & $\dfrac{\varepsilon}{k}$ & $\hat{\textbf{S}}$ \\ [1ex]% inserting body of the table
$\mathbf{\Omega}$ & $\left| \mathbf{\Omega} \right|$ & $\hat{\mathbf{\Omega}}$ \\ [1ex]
$\nabla \overline{p}$ & $\rho \left| \frac{D\overline{\textbf{u}}}{Dt} \right|$ & $\widehat{\nabla \overline{p}}$ \\ [1ex]
$\nabla k$ & $\dfrac{\varepsilon}{\sqrt{k}}$ & $\widehat{\nabla k}$ \\ [1ex]
$k$ & $\nu |\textbf{S}|$ & $\hat{k}$ \\ [1ex]
$T_t$ & $|\textbf{S}|^{-1}$ & $\hat{T}_t$ \\ [1ex]
$\nu_t$ & $\nu$ & $\hat{\nu}_t$ \\ [1ex]
\hline
\end{tabular}
\label{table:nondimensional_inputs}
\end{table}

\begin{table}[t!]
\caption{Nondimensional frame invariant inputs for statistically three-dimensional turbulent flows.  Nondimensional frame invariant inputs for statistically two-dimensional flows are highlighted in blue.  Below, $\ph_{lm} = \epsilon_{lmn} ( \widehat{\nabla \overline{p}} )_n$ and $\kh_{lm} = \epsilon_{lmn} ( \widehat{\nabla k} )_n$.}
%\centering % used for centering table
\begin{adjustbox}{center}
\begin{tabular}{| c | c | c | c | c | c |} % centered columns (4 columns)
\hline &&&&& \\[-1em]
Input & Relation & Input & Relation &Input & Relation  \\ [0.5ex] % inserts table
\hline &&&&& \\[-1em]
{\color{blue} $\hat{q}_{1}$} & {\color{blue}$\sh_{ij} \sh_{ji}$} & $\hat{q}_{17}$ &$\oh_{ij} \ph_{jk} \sh_{ki}  $ &$\hat{q}_{33}$ &$\kh_{ij} \kh_{jk} \sh_{kl} \oh_{lm} \sh_{mn} \sh_{ni}   $ \\ [1ex]
$\hat{q}_{2}$ &$\sh_{ij}\sh_{jk}\sh_{ki}$ & $\hat{q}_{18}$ &$\oh_{ij} \kh_{ji}$ & {\color{blue}$\hat{q}_{34}$} & {\color{blue}$\ph_{ij} \kh_{jk} \sh_{ki}$} \\ [1ex]
{\color{blue} $\hat{q}_{3}$} & {\color{blue}$\oh_{ij}\oh_{ji}$} & $\hat{q}_{19}$ &$\oh_{ij} \ph_{jk} \sh_{kl} \sh_{li}    $ &$\hat{q}_{35}$ &$\ph_{ij} \kh_{jk} \sh_{kl} \sh_{li} $ \\ [1ex]
{\color{blue} $\hat{q}_{4}$} & {\color{blue}$\ph_{ij}\ph_{ji}$} & $\hat{q}_{20}$ &$\oh_{ij} \oh_{jk} \ph_{kl} \sh_{li}    $ &$\hat{q}_{36}$ &$\ph_{ij} \ph_{jk} \kh_{kl} \sh_{li}  $ \\ [1ex]
{\color{blue} $\hat{q}_{5}$} & {\color{blue}$\kh_{ij}\kh_{ji}$} & {\color{blue}$\hat{q}_{21}$} & {\color{blue}$\ph_{ij} \ph_{jk} \oh_{kl} \sh_{li} $} & $\hat{q}_{37}$ &$\kh_{ij} \kh_{jk} \ph_{kl} \sh_{li}$ \\ [1ex]
$\hat{q}_{6}$ & $\oh_{ij} \oh_{jk} \sh_{ki} $& $\hat{q}_{22}$ &$\oh_{ij} \oh_{jk} \ph_{kl} \sh_{lm} \sh_{mi}   $   &$\hat{q}_{38}$ & $\ph_{ij} \ph_{jk} \kh_{kl} \sh_{lm} \sh_{mi} $ \\ [1ex]
$\hat{q}_{7}$ &$\oh_{ij} \oh_{jk} \sh_{kl} \sh_{li} $ & $\hat{q}_{23}$ &$\ph_{ij} \ph_{jk} \oh_{kl} \sh_{lm} \sh_{mi}$ &$\hat{q}_{39}$ &$\kh_{ij} \kh_{jk} \ph_{kl} \sh_{lm} \sh_{mi} $ \\ [1ex]
$\hat{q}_{8}$ &$\oh_{ij} \oh_{jk} \sh_{kl} \oh_{lm} \sh_{mn} \sh_{ni} $ & $\hat{q}_{24}$ &$\oh_{ij} \oh_{jk} \sh_{kl} \ph_{lm} \sh_{mn} \sh_{ni}  $ &$\hat{q}_{40}$ &$\ph_{ij} \ph_{jk} \sh_{kl} \kh_{lm} \sh_{mn} \sh_{ni}$ \\ [1ex]
{\color{blue} $\hat{q}_{9}$} & {\color{blue}$\ph_{ij} \ph_{jk} \sh_{ki}$} & $\hat{q}_{25}$ &$\ph_{ij} \ph_{jk} \sh_{kl} \oh_{lm} \sh_{mn} \sh_{ni} $ & $\hat{q}_{41}$ &$ \kh_{ij} \kh_{jk} \sh_{kl} \ph_{lm} \sh_{mn} \sh_{ni} $ \\ [1ex]
$\hat{q}_{10}$ &$\ph_{ij} \ph_{jk} \sh_{kl} \sh_{li}  $ &$\hat{q}_{26}$ &$\oh_{ij} \kh_{jk} \sh_{ki} $ & {\color{blue}$\hat{q}_{42}$} & {\color{blue}$\oh_{ij} \ph_{jk} \kh_{ki}$} \\ [1ex]
$\hat{q}_{11}$ &$\ph_{ij} \ph_{jk}  \sh_{kl}  \ph_{lm}   \sh_{mn}  \sh_{ni}    $ &$\hat{q}_{27}$ &$\oh_{ij} \kh_{jk} \sh_{kl} \sh_{li}  $	 & {\color{blue} $\hat{q}_{43}$} & {\color{blue}$\oh_{ij} \ph_{jk} \kh_{kl} \sh_{li}$} \\ [1ex]
{\color{blue} $\hat{q}_{12}$} & {\color{blue}$\kh_{ij} \kh_{jk} \sh_{ki}$} & $\hat{q}_{28}$ &$\oh_{ij} \oh_{jk} \kh_{kl} \sh_{li}$ & $\hat{q}_{44}$ &$ \oh_{ij} \kh_{jk} \ph_{kl} \sh_{li} $ \\ [1ex]
$\hat{q}_{13}$ &$\kh_{ij} \kh_{jk} \sh_{kl} \sh_{li} $ & {\color{blue}$\hat{q}_{29}$} & {\color{blue}$\kh_{ij} \kh_{jk} \oh_{kl} \sh_{li}$} & $\hat{q}_{45}$ &$\oh_{ij} \ph_{jk} \kh_{kl} \sh_{lm} \sh_{mi}$ \\ [1ex]
$\hat{q}_{14}$ &$\kh_{ij} \kh_{jk} \sh_{kl} \kh_{lm} \sh_{mn} \sh_{ni} $ &$\hat{q}_{30}$ &$\oh_{ij} \oh_{jk} \kh_{kl} \sh_{lm} \sh_{mi} $ &$\hat{q}_{46}$ & $\oh_{ij} \kh_{jk} \ph_{kl} \sh_{lm} \sh_{mi}$ \\ [1ex]
$\hat{q}_{15}$ &$\oh_{ij} \ph_{ji}   $ &$\hat{q}_{31}$ &$\kh_{ij} \kh_{jk} \oh_{kl} \sh_{lm} \sh_{mi}$ &$\hat{q}_{47}$ &$ \oh_{ij} \ph_{jk} \sh_{kl} \kh_{lm} \sh_{mn} \sh_{ni}$ \\ [1ex]
{\color{blue} $\hat{q}_{16}$} & {\color{blue}$\ph_{ij} \kh_{ji}$} & $\hat{q}_{32}$ &$\oh_{ij} \oh_{jk} \sh_{kl} \kh_{lm} \sh_{mn} \sh_{ni}   $ & {\color{blue} $\hat{q}_{48}$, $\hat{q}_{49}$, $\hat{q}_{50}$} & {\color{blue}$\hat{k}$, $\hat{T}_t$, $\hat{\nu}_t$} \\ [1 ex]
\hline %inserts single line
\end{tabular}
\end{adjustbox}
\label{table:invariant_inputs} % is used to refer this table in the text
\end{table}

\subsection{Frame Invariance}
\label{subsec:frameinvariance}
To arrive at a \textit{frame invariant} $S$-frame discrepancy model, that is, a model that maps as a tensor under a change of coordinates, the inputs of the model must also be frame invariant.  Unfortunately, the inputs to \eqref{eq:nondimensional_model_form} are \textit{not} frame invariant.  Fortunately, Hilbert's basis theorem may be employed to arrive at models which are frame invariant \cite{hilbert1993theory}.  The principal idea is that any frame invariant tensor-valued function of a set of tensors may be written in terms of \textit{polynomial invariants} of the set of tensors.  While sets of tensors generally have an infinite number of polynomial invariants, Hilbert's basis theorem states that each polynomial invariant may be expressed as a polynomial of the components of a finite-dimensional \textit{minimal integrity basis}.  Application of Hilbert's basis theorem to \eqref{eq:nondimensional_model_form} yields frame invariant nondimensional $S$-frame discrepancy models of the form
\begin{equation}
\hat{D}^{S}_{ij} = \hat{D}^{S,\text{inv-model}}_{ij}\left( \hat{\textbf{q}} \right),
\label{eq:invariant_model_form}
\end{equation}
where $\hat{\textbf{q}}$ is a nondimensional input vector whose components belong to the minimal integrity basis for the nondimensional set of tensors
\begin{equation}
\hat{\mathcal{Q}} = \left\{ \hat{\textbf{S}}, \hat{\mathbf{\Omega}}, \widehat{\nabla \overline{p}}, \widehat{\nabla k}, \hat{k}, \hat{T}_t, \hat{\nu}_t \right\}.
\end{equation}
The precise components of $\hat{\textbf{q}}$ are listed in Table \ref{table:invariant_inputs}, wherein $\ph_{lm} = \epsilon_{lmn} ( \widehat{\nabla \overline{p}} )_n$ and $\kh_{lm} = \epsilon_{lmn} ( \widehat{\nabla k} )_n$.  There are fifty nondimensional frame invariant inputs in the statistically three-dimensional setting and fifteen nondimensional frame invariant inputs in the statistically two-dimensional setting.

\subsection{Functional Mapping using Artificial Neural Networks}
\label{subsec:neuralnets}
To arrive at a \textit{computable} $S$-frame discrepancy model, we employ an \textit{articial neural network}  (ANN) to represent each of the functions $\hat{D}^{S,\text{inv-model}}_{ij}$ appearing in \eqref{eq:invariant_model_form}.  In particular, we employ \textit{dense multi-layer feed-forward neural networks}.  In previous work, ANN discrepancy models have been shown to be both accurate and computationally efficient for challenging flows \cite{tracey2015machine,ling2016reynolds}, and we have found that ANNs typically yield improved performance over alternative regression techniques such as random forests \cite{wang2017physics} and polynomial expansions \cite{schmelzer2020discovery} in modeling the components of the discrepancy tensor in the $S$-frame on an accuracy-versus-cost basis.  For the sake of brevity, we do not review the construction of deep neural networks in this paper, and instead we refer the reader to \cite{goodfellow2016deep,higham2019deep} for more details.  All of the flows analyzed in this paper are statistically two-dimensional, so the ANN models we construct in this paper are functions only of the inputs highlighted in blue in Table \ref{table:invariant_inputs}.  However, one may also construct ANN models for statistically three-dimensional flows using the procedure outlined in this section.

\subsection{Learning Procedure and Hyperparameter Selection}
\label{subsec:learning}
To learn the weights and biases in our ANN $S$-frame discrepancy model for a particular ANN architecture, we employ a \textit{mean squared error (MSE) loss function} of the form
\begin{equation}
    \text{MSE}_\text{train}(\hat{\textbf{W}},\hat{\textbf{b}}) = \frac{1}{n_\text{train}} \sum_{a = 1}^{n_\text{train}} \left\| \hat{\textbf{D}}^\text{DNS-train}_a - \hat{\textbf{D}}^{\text{ANN}}\left(\hat{\textbf{q}}_a^\text{DNS-train}; \hat{\textbf{W}},\hat{\textbf{b}}\right) \right\|_F^2,
    \label{eq:mse_train}
\end{equation}
where $\hat{\textbf{D}}^\text{\text{ANN}}$ denotes the nondimensional tensor form of our ANN $S$-frame discrepancy model, $\hat{\textbf{W}}$ and $\hat{\textbf{b}}$ denote the weights and biases of our ANN model, $\left\{ \hat{\textbf{D}}^\text{DNS-train}_a \right\}_{a=1}^{n_\text{train}}$ denote the values of the nondimensional discrepancy tensor as obtained by DNS at a set of \textit{training} points, $\left\{ \hat{\textbf{q}}^\text{DNS-train}_a \right\}_{a=1}^{n_\text{train}}$ denote the values of the nondimensional input vector as obtained by DNS at the same set of training points, and $\| \cdot \|_F$ denotes the Frobenius norm.  We utilize \textit{Adam}, an adaptive learning rate optimization algorithm designed specifically for training deep neural networks, in order to find the optimal weights and biases associated with the above loss function \cite{kingma2014adam}.  MSE loss functions are the most common type of loss function in training ANNs as they are simple and differentiable, though they tend to overemphasize individual large errors \cite{geman1992neural}.  Other types of cost function may also be employed to avoid this issue, but we have found that MSE loss functions yield ANN $S$-frame discrepancy models that are sufficiently accurate and robust for canonical turbulent flow problems.

To optimize our ANN $S$-frame discrepancy model architecture (i.e., the number of layers, the number of neurons per layer, and the activation function for each neuron), we first evaluate the performance of an optimized ANN $S$-frame discrepancy model for a given architecture using the \textit{mean relative error (MRE)}
\begin{equation}
    \text{MRE}_\text{validate}(\hat{\textbf{W}},\hat{\textbf{b}}) = \frac{1}{n_\text{validate}} \sum_{a = 1}^{n_\text{validate}} \frac{ \left\| \hat{\textbf{D}}^\text{DNS-validate}_a - \hat{\textbf{D}}^{\text{ANN}}\left(\hat{\textbf{q}}_a^\text{DNS-validate}; \hat{\textbf{W}},\hat{\textbf{b}}\right) \right\|_F }{\left\| \hat{\textbf{D}}^\text{DNS-validate}_a \right\|_F}
    \label{eq:mre_validate}
\end{equation}
evaluated at a set of \textit{validation} points which does not overlap with the training dataset.  If the MRE is found to be too large, we then attempt to improve our ANN $S$-frame discrepancy model by increasing the number of layers, increasing the number of neurons per layer, or selecting different activation functions.  As the computational cost associated with an ANN model increases with an increasing number of layers or neurons per layer, we seek the smallest architecture (i.e., the architecture with the least number of weights and biases) which delivers acceptable model performance.  Finally, to assess the performance of our final optimized ANN $S$-frame discrepancy model, we use the MRE
\begin{equation}
    \text{MRE}_\text{test}(\hat{\textbf{W}},\hat{\textbf{b}}) = \frac{1}{n_\text{test}} \sum_{a = 1}^{n_\text{test}} \frac{ \left\| \hat{\textbf{D}}^\text{DNS-test}_a - \hat{\textbf{D}}^{\text{ANN}}\left(\hat{\textbf{q}}_a^\text{DNS-test}; \hat{\textbf{W}},\hat{\textbf{b}}\right) \right\|_F }{\left\| \hat{\textbf{D}}^\text{DNS-test}_a \right\|_F}
    \label{eq:mre_test}
\end{equation}
evaluated at a set of \textit{testing} points which does not overlap with either the training or validation datasets.  We also examine the MRE for individual components of the discrepancy tensor in the $S$-frame to assess how well the final optimized ANN $S$-frame discrepancy model predicts each component of the discrepancy tensor.

In this paper, we select the training, validation, and testing datasets from the same DNS databases, and we consider a 50-25-25 split across the training, validation, and testing datasets.  Moreover, we build two classes of ANN $S$-frame discrepancy models.  For the first class of models, which we denote as \textit{specific} models, we employ data obtained from a single DNS simulation for training, validation, and testing.  For the second class of models, which we denote as \textit{universal} models, we employ data associated from several DNS simulations for training, validation, and testing.  We expect specific models to outperform universal models for the specific test case they were trained on, but we do not expect specific models to perform well for other test cases.  Since universal models are trained on several test cases, we expect them to better generalize than specific models.

\section{Energetic Stability}
\label{sec:energy}
A given Reynolds stress closure is practically useless if it yields an \textit{unstable} RANS model, even if the Reynolds stress closure is highly accurate \cite{wu2019reynolds}.  Thus, in this section, we study the stability properties of $S$-frame discrepancy models, and we discuss a procedure for direct enforcement of \textit{global decay} of \textit{mean kinetic energy}.  To begin, we show that any $S$-frame discrepancy model admits a \textit{balance law} for mean kinetic energy.\\
\\
\textbf{Proposition 1} (Balance of Mean Kinetic Energy for $S$-frame Discrepancy Models)
\textit{Let $\Omega \subset \mathbb{R}^3$ be a domain of interest.  If no-slip boundary conditions are enforced along the domain boundary (i.e., if $\overline{u}_i|_{\partial \Omega} = 0$) and no body forces are applied within the domain (i.e., if $\overline{f}_i|_{\Omega} = 0$), then a dimensional $S$-frame discrepancy model of the form given by \eqref{eq:dimensional_model_form} admits the mean kinetic energy balance law
\begin{equation}
\frac{d}{dt} \int_{\Omega} \frac{1}{2} {\overline{u}_i}^2 d\Omega = - \int_{\Omega} \left( 2(\nu + \nu_t) S^2 - D_{11}^{S,\textup{model}} \lambda^S_{1} - D_{22}^{S,\textup{model}} \lambda^S_{2}- D_{33}^{S,\textup{model}} \lambda^S_{3} \right) d\Omega,
\label{energy0}
\end{equation}
where ${\overline{u}_i}^2 = {\overline{u}_i}{\overline{u}_i}$, $S^2 = S_{ij}S_{ij}$, and $\left\{ \lambda^S_i \right\}_{i=1}^3$ are the eigenvalues of the mean strain-rate tensor.}\\
\\
\textit{Proof}.  The proof begins by contracting \eqref{NS} with $\overline{u}_i$ and integrating over the domain $\Omega$, resulting in
\begin{equation}
\int_{\Omega} \left( \overline{u}_i \overline{u}_{i,t} + \overline{u}_i \overline{u}_j \overline{u}_{i,j} + \overline{u}_i \frac{1}{\rho} \overline{p}_{,i} - \overline{u}_i \left(  2\nu S_{ij}\right)_{,j} + \overline{u}_i \tau_{ij,j} \right) d\Omega = \int_{\Omega}  \overline{u}_i \overline{f}_i d\Omega.
\end{equation}
The Reynolds stress tensor admits the decomposition $\tau_{ij} = -2\nu_t S_{ij} + \frac{2}{3} k \delta_{ij} + D_{ij}$.  Moreover, for the case in consideration, $\overline{f}_i = 0$.  It then follows that
\begin{equation}
\int_{\Omega} \left( \overline{u}_i \overline{u}_{i,t} + \overline{u}_i \overline{u}_j \overline{u}_{i,j} + \overline{u}_i \left( \frac{\overline{p}}{\rho} + \frac{2}{3} k \right)_{,i} - \overline{u}_i \left(  2(\nu + \nu_t) S_{ij}\right)_{,j} + \overline{u}_i D_{ij,j} \right) d\Omega = 0.
\end{equation}
By the product rule,
\begin{equation}
\frac{d}{dt} \int_{\Omega} \frac{1}{2} \overline{u}^2_i d\Omega + \int_{\Omega} \left( \overline{u}_j \left( \frac{1}{2} \overline{u}^2_{i} \right)_{,j} + \overline{u}_i \left( \frac{\overline{p}}{\rho} + \frac{2}{3} k \right)_{,i} - \overline{u}_i \left(  2(\nu + \nu_t) S_{ij}\right)_{,j} + \overline{u}_i D_{ij,j} \right) d\Omega = 0.
\end{equation}
By integration by parts and the fact that $\overline{u}_i|_{\partial \Omega} = 0$ and $\overline{u}_{i,i} = \overline{u}_{j,j} = 0$,
\begin{equation}
\frac{d}{dt} \int_{\Omega} \frac{1}{2} \overline{u}^2_i d\Omega + \int_{\Omega} \overline{u}_{i,j} \left(  2(\nu + \nu_t) S_{ij} - D_{ij} \right) d\Omega = 0.
\end{equation}
Since the mean strain-rate tensor and discrepancy tensor are symmetric,
\begin{equation}
\frac{d}{dt} \int_{\Omega} \frac{1}{2} \overline{u}^2_i d\Omega + \int_{\Omega} S_{ij} \left(  2(\nu + \nu_t) S_{ij} - D_{ij} \right) d\Omega = 0.
\end{equation}
The desired result then follows by recognizing $S_{ij} S_{ij} = S^2$, $S_{ij} D_{ij} = D_{11}^{S} \lambda^S_{1} + D_{22}^{S} \lambda^S_{2} + D_{33}^{S} \lambda^S_{3}$, and $D^S_{ij} = D^{S,\text{model}}_{ij}\left(\textbf{S}, \mathbf{\Omega}, \nabla \overline{p}, \nabla k, k, T_t, \nu_t\right)$. $\qed$\\
\\
From the above result, we infer that a given $S$-frame discrepancy model results in a global decay of mean kinetic energy if and only if
\begin{equation}
\int_{\Omega} \left( 2(\nu + \nu_t) S^2 - D_{11}^{S,\textup{model}} \lambda^S_{1} - D_{22}^{S,\textup{model}} \lambda^S_{2}- D_{33}^{S,\textup{model}} \lambda^S_{3} \right) d\Omega \geq 0. \label{eq:global_constraint}
\end{equation}
While it cannot be proven that there is a global decay of mean kinetic energy for general turbulent flow problems, a RANS model may exhibit blow up if it allows for accretion of mean kinetic energy.  To enforce a global decay of mean kinetic energy, we take advantage of the fact that the \textit{global} inequality \eqref{eq:global_constraint} holds if the \textit{local} inequality
\begin{equation}
D_{11}^{S,\textup{model}} \lambda^S_{1} + D_{22}^{S,\textup{model}} \lambda^S_{2} + D_{33}^{S,\textup{model}} \lambda^S_{3} \leq 2(\nu + \nu_t) S^2 \label{eq:local_constraint}
\end{equation}
also holds throughout the domain.  While \eqref{eq:local_constraint} is a stronger condition than \eqref{eq:global_constraint}, it is possible to construct a discrepancy model satisfying \eqref{eq:local_constraint} that depends only on local values of mean flow and turbulence variables.  In particular, given an $S$-frame discrepancy model of the form
\begin{equation}
D^S_{ij} = D^{S,\text{model}}_{ij}\left( \textbf{s} \right),
\end{equation}
where $\textbf{s} = \left( \textbf{S}, \mathbf{\Omega}, \nabla \overline{p}, \nabla k, k, T_t, \nu_t \right)$, a corresponding discrepancy model satisfying \eqref{eq:local_constraint} is
\begin{equation}
\textbf{D}^{\text{stable-model}}\left( \textbf{s} \right) = \argmin_{\tilde{\textbf{D}} \in \mathcal{S}(\textbf{s})} \frac{1}{2} \sum_{i,j} \left( \tilde{D}_{ij}^S- D^{S,\text{model}}_{ij}\left( \textbf{s} \right) \right)^2, \label{eq:stable_model}
\end{equation}
where
\begin{equation}
\mathcal{S}(\textbf{s}) := \left\{ \tilde{\textbf{D}} \in \text{Sym}_0(3) : \tilde{D}_{11}^{S} \lambda^S_{1} + \tilde{D}_{22}^{S} \lambda^S_{2} + \tilde{D}_{33}^{S} \lambda^S_{3} - 2(\nu + \nu_t) S^2 \leq 0 \right\}
\end{equation}
and $\text{Sym}_0(3)$ is the space of symmetric traceless rank-2 tensors in $\mathbb{R}^3$.   Note if the original $S$-frame discrepancy model is Galilean invariant and frame invariant, so is the post-processed discrepancy model.

The minimization problem given by \eqref{eq:stable_model} is a linearly-constrained quadratic program with strictly convex objective functional and nonempty feasible set.  Thus, it has a unique solution which may be obtained using state-of-the-art quadratic programming methods (see, e.g., \cite[Chapter~16]{nocedal2006numerical}).  For instance, the active set method may be employed to solve the associated Karush-Kuhn-Tucker optimality conditions.  If
\begin{equation}
D_{11}^{S,\textup{model}}\left( \textbf{s} \right) \lambda^S_{1} + D_{22}^{S,\textup{model}}\left( \textbf{s} \right) \lambda^S_{2} + D_{33}^{S,\textup{model}}\left( \textbf{s} \right) \lambda^S_{3} \leq 2(\nu + \nu_t) S^2,
\end{equation}
then the active set method returns $D_{ij}^{S,\text{stable-model}}\left( \textbf{s} \right) = D_{ij}^{S,\textup{model}}\left( \textbf{s} \right)$.  Otherwise, the components $D_{ij}^{S,\text{stable-model}}\left( \textbf{s} \right)$ are obtained via solution of the linear system
\begin{equation}
\left[
\begin{array}{cccccccc}
1 & 0 & 0 & 0 & 0 & 0 & \lambda^S_1 & 1 \\
0 & 1 & 0 & 0 & 0 & 0 & \lambda^S_2 & 1 \\
0 & 0 & 1 & 0 & 0 & 0 & \lambda^S_3 & 1 \\
0 & 0 & 0 & 1 & 0 & 0 & 0 & 0 \\
0 & 0 & 0 & 0 & 1 & 0 & 0 & 0 \\
0 & 0 & 0 & 0 & 0 & 1 & 0 & 0 \\
\lambda^S_1 & \lambda^S_2 & \lambda^S_3 & 0 & 0 & 0 & 0 & 0 \\
1 & 1 & 1 & 0 & 0 & 0 & 0 & 0
\end{array}
\right] \left[
\begin{array}{c}
D_{11}^{S,\textup{stable-model}}\left( \textbf{s} \right) \\
D_{22}^{S,\textup{stable-model}}\left( \textbf{s} \right) \\
D_{33}^{S,\textup{stable-model}}\left( \textbf{s} \right) \\
D_{12}^{S,\textup{stable-model}}\left( \textbf{s} \right) \\
D_{13}^{S,\textup{stable-model}}\left( \textbf{s} \right) \\
D_{23}^{S,\textup{stable-model}}\left( \textbf{s} \right) \\
\mu_1\left( \textbf{s} \right) \\
\mu_2\left( \textbf{s} \right)
\end{array}
\right] = \left[
\begin{array}{c}
D_{11}^{S,\textup{model}}\left( \textbf{s} \right) \\
D_{22}^{S,\textup{model}}\left( \textbf{s} \right) \\
D_{33}^{S,\textup{model}}\left( \textbf{s} \right) \\
D_{12}^{S,\textup{model}}\left( \textbf{s} \right) \\
D_{13}^{S,\textup{model}}\left( \textbf{s} \right) \\
D_{23}^{S,\textup{model}}\left( \textbf{s} \right) \\
2(\nu + \nu_t)S^2 \\
0
\end{array}
\right], \label{eq:linear_sys}
\end{equation}
where $\mu_1\left( \textbf{s} \right)$ is the Lagrange multiplier associated with the local inequality constraint \eqref{eq:local_constraint} and $\mu_2\left( \textbf{s} \right)$ is the Lagrange multiplier associated with the traceless constraint on $\textbf{D}^{\text{stable-model}}\left( \textbf{s} \right)$.

\section{Realizability}
\label{sec:realizability}
As previously mentioned in this paper, $S$-frame discrepancy models are not necessarily \textit{realizable}.  That is, the eigenvalues of the Reynolds stress tensor as predicted by an $S$-frame discrepancy model are not guaranteed to be non-negative.  However, like energetic stability, realizability may be attained via a post-processing procedure.  In particular, given an $S$-frame discrepancy model of the form
\begin{equation}
D^S_{ij} = D^{S,\text{model}}_{ij}\left( \textbf{s} \right),
\end{equation}
where $\textbf{s} = \left( \textbf{S}, \mathbf{\Omega}, \nabla \overline{p}, \nabla k, k, T_t, \nu_t \right)$, a corresponding realizable discrepancy model is
\begin{equation}
\textbf{D}^{\text{realizable-model}}\left( \textbf{s} \right) = \argmin_{\tilde{\textbf{D}} \in \mathcal{R}(\textbf{s})} \frac{1}{2} \sum_{i,j} \left( \tilde{D}_{ij}^S- D^{S,\text{model}}_{ij}\left( \textbf{s} \right) \right)^2, \label{eq:realizable_model}
\end{equation}
where
\begin{equation}
\mathcal{R}(\textbf{s}) := \left\{ \tilde{\textbf{D}} \in \text{Sym}_0(3) : 
\begin{array}{c}
-2 \nu_t \lambda^S_1 + \tilde{D}_{11}^{S} + \frac{2}{3} k \geq 0\\
-2 \nu_t \lambda^S_2 + \tilde{D}_{22}^{S} + \frac{2}{3} k \geq 0\\
-2 \nu_t \lambda^S_3 + \tilde{D}_{33}^{S} + \frac{2}{3} k \geq 0\\
\tilde{D}_{12}^{S} \leq \left(-2 \nu_t \lambda^S_1 + \tilde{D}_{11}^{S} + \frac{2}{3} k\right) \left(-2 \nu_t \lambda^S_2 + \tilde{D}_{22}^{S} + \frac{2}{3} k \right) \\
\tilde{D}_{13}^{S} \leq \left(-2 \nu_t \lambda^S_1 + \tilde{D}_{11}^{S} + \frac{2}{3} k\right) \left(-2 \nu_t \lambda^S_3 + \tilde{D}_{33}^{S} + \frac{2}{3} k \right) \\
\tilde{D}_{23}^{S} \leq \left(-2 \nu_t \lambda^S_2 + \tilde{D}_{22}^{S} + \frac{2}{3} k\right) \left(-2 \nu_t \lambda^S_3 + \tilde{D}_{33}^{S} + \frac{2}{3} k \right) \\
-\tilde{D}_{12}^{S} \leq \left(-2 \nu_t \lambda^S_1 + \tilde{D}_{11}^{S} + \frac{2}{3} k\right) \left(-2 \nu_t \lambda^S_2 + \tilde{D}_{22}^{S} + \frac{2}{3} k \right) \\
-\tilde{D}_{13}^{S} \leq \left(-2 \nu_t \lambda^S_1 + \tilde{D}_{11}^{S} + \frac{2}{3} k\right) \left(-2 \nu_t \lambda^S_3 + \tilde{D}_{33}^{S} + \frac{2}{3} k \right) \\
-\tilde{D}_{23}^{S} \leq \left(-2 \nu_t \lambda^S_2 + \tilde{D}_{22}^{S} + \frac{2}{3} k\right) \left(-2 \nu_t \lambda^S_3 + \tilde{D}_{33}^{S} + \frac{2}{3} k \right) \\
\end{array}\right\}.
\end{equation}
The minimization problem given by \eqref{eq:stable_model} is a quadratically-constrained quadratic program.  Quadratically-constrained quadratic programs are much more difficult to solve than linearly-constrained quadratic programs, and in fact, quadratically-constrained quadratic programs are generally NP-hard \cite{anstreicher2009semidefinite}.  Nonetheless, state-of-the-art nonlinear programming methods may be employed to solve \eqref{eq:stable_model} (see, e.g., \cite[Chapters~17-18]{nocedal2006numerical}).  It should also be noted an energetically stable and realizable model can be constructed from the original $S$-frame discrepancy model in similar fashion, i.e.,
\begin{equation}
\textbf{D}^{\text{stable-and-realizable-model}}\left( \textbf{s} \right) = \argmin_{\tilde{\textbf{D}} \in \mathcal{S}(\textbf{s}) \cap \mathcal{R}(\textbf{s})} \frac{1}{2} \sum_{i,j} \left( \tilde{D}_{ij}^S- D^{S,\text{model}}_{ij}\left( \textbf{s} \right) \right)^2. \label{eq:stable_and_realizable_model}
\end{equation}
However, realizability is not directly enforced in any of the numerical experiments appearing in this paper, and we have found that enforcing realizability is not a requirement for attaining stable and accurate results with an $S$-frame discrepancy model.

\section{Numerical Results}
\label{sec:numerical}
Now that we have discussed how to construct $S$-frame discrepancy models using high-fidelity simulation data, we are ready to test the accuracy of such Reynolds stress closures.  In this section, we present numerical results using both specific and universal ANN $S$-frame discrepancy models and four turbulent flow configurations for which DNS data is available.  In particular, we examine the accuracy of $S$-frame discrepancy models in predicting the Reynolds stress tensor, and we propagate $S$-frame discrepancy models through a flow solver to determine their accuracy in predicting the mean flow field.

\subsection{Selected Test Cases}
\label{subsec:testdata}
To test the effectiveness of $S$-frame discrepancy models for data-informed Reynolds stress closure, four turbulent flow configurations are considered in this paper.  Each of the selected flow configurations is statically stationary and two-dimensional, and DNS data is available for each configuration.  The selected flow configurations also cover a wide range of turbulent flow regimes of engineering interest, from fully attached turbulent flow to turbulent flow exhibiting either shallow or massive flow separation.  We know of no existing RANS model which is predictive for each flow configuration.

The first flow configuration considered in this paper is turbulent channel flow.  Turbulent channel flow is a canonical test problem for assessing the accuracy of RANS models.  Of particular interest is a RANS model's ability to replicate the so-called law of the wall \cite{von1931mechanical}.  We consider turbulent channel flow at a friction Reynolds number of $Re_{\tau} = 550$, for which high quality DNS data is available \cite{lee2015direct}.  For the simulations reported in this paper, the channel is selected to have dimensions 0.5 and 1.0 in the stream-wise and wall-normal directions, respectively.  Due to symmetry, only half the channel is considered.  Periodic boundary conditions are applied in the stream-wise direction, no-slip boundary conditions are applied along the bottom wall, and a symmetry condition is applied along the centerline of the channel.  The density and kinematic viscosity are selected to be $\rho = 1$ and $\nu =$ 1e-4, respectively.  An external pressure gradient is imposed in the stream-wise direction, and the value of this gradient is selected to ensure the correct friction Reynolds number is achieved.  A sufficiently fine mesh is employed to fully resolve the mean velocity and pressure fields, and a non-uniform grid spacing is utilized in the wall-normal direction to fully resolve the boundary layer.

\begin{figure}[t!]
  \centering
  \includegraphics[scale=.25]{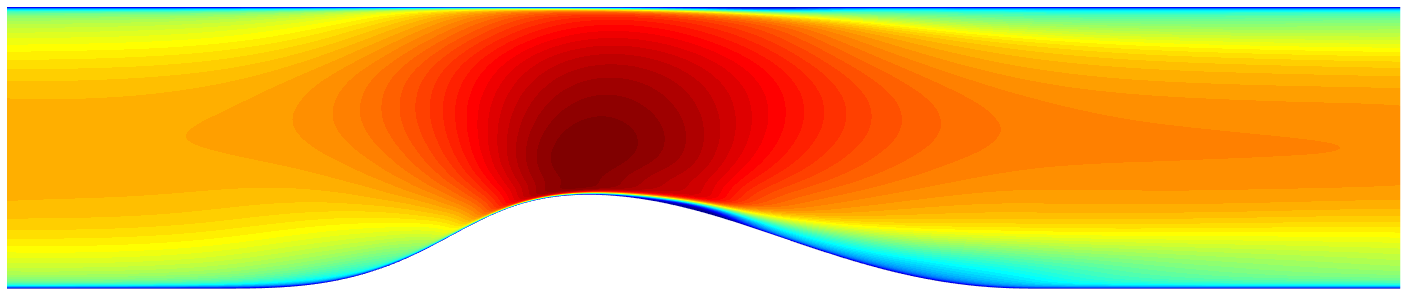}
\caption{The mean $x$-velocity profile for the two-dimensional bump test case.}
\label{Figure8}
\end{figure}

The second flow configuration considered in this paper is turbulent flow over a two-dimensional bump.  DNS data was recently obtained for this flow configuration for an inflow friction Reynolds number of $Re_{\tau} = 600$ \cite{marquillie2008direct}.  The mean $x$-velocity profile computed from this DNS data is displayed in Figure \ref{Figure8}.  For this flow configuration, shallow flow separation is induced by an adverse pressure gradient which directly follows a favorable pressure gradient due to a smooth change in geometry.  Even though the separation region is relatively small, our numerical experiments indicate that two commonly employed LEV models, namely the Spalart-Allmaras \cite{spalart1992one} and Menter SST \cite{menter1994two} models, fail to yield accurate predictions of the separation and reattachment points for this flow configuration.  For the simulations reported in this paper, no-slip boundary conditions are applied along the top and bottom walls.  A zero-traction boundary condition is set at the outlet, and mean velocity boundary conditions are set at the inlet by interpolating DNS data.  The density and kinematic viscosity are selected to be $\rho = 1$ and $\nu =$ 7.9365e-5, respectively.  No body forces are applied.  As is the case for the channel flow simulations, a sufficiently fine mesh is employed to fully resolve the mean velocity and pressure fields.

\begin{figure}[t!]
  \centering
  \includegraphics[scale=.26]{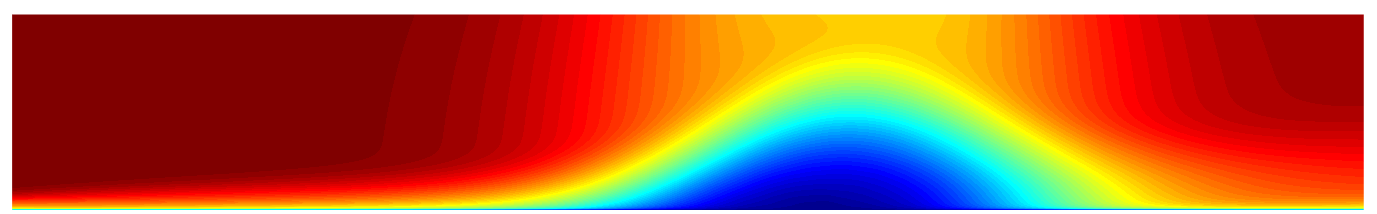}
\caption{The mean $x$-velocity profile for the separation bubble test case.}
\label{Figure9}
\end{figure}

\begin{figure}[t!]
  \centering
  \includegraphics[scale=.225]{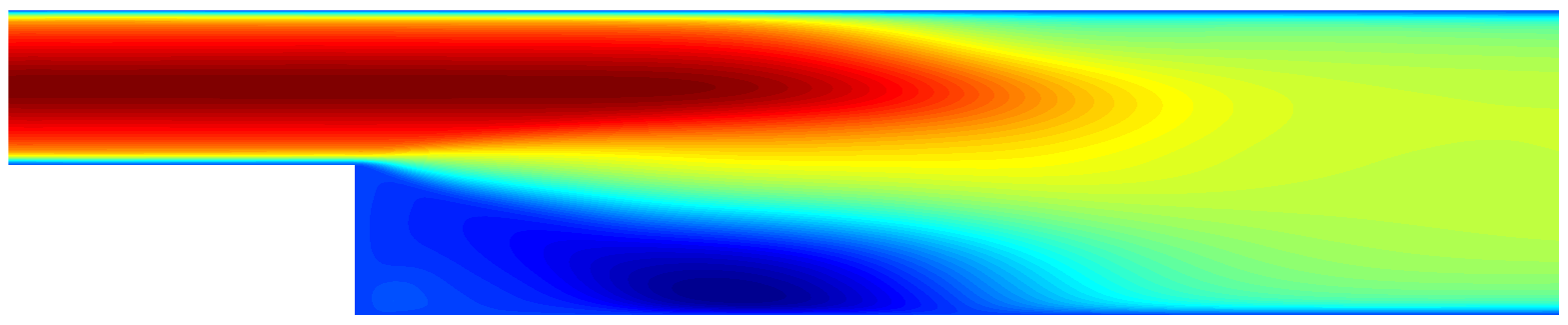}
\caption{The mean $x$-velocity profile for the backward facing step test case.}
\label{Figure10}
\end{figure}

The third flow configuration considered in this paper is a turbulent separation bubble produced in a channel by suction and blowing along the top wall.  DNS data was recently obtained for this flow configuration for an inflow momentum Reynolds number of $Re_{\theta} = 2000$ \cite{coleman2018numerical}.  The mean $x$-velocity profile computed from this DNS data is displayed in Figure \ref{Figure9}.  For this flow configuration, separation is not caused by the presence of an adverse pressure gradient due to curvature but instead by the suction and blowing.  The results of \cite{coleman2018numerical} demonstrate several state-of-the-art LEV models struggle to accurately predict the skin friction coefficient in the region of the turbulent separation bubble, and consequently, they fail to accurately predict the reattachment length.  For the simulations reported in this paper, no-slip boundary conditions are applied along the bottom wall.  A zero-traction boundary condition is set at the outlet, and mean velocity boundary conditions are set at the inlet by interpolating DNS data.  Traction boundary conditions computed from the DNS data are applied along the top wall to yield the desired effect of suction and blowing.  The density and kinematic viscosity are selected to be $\rho = 1$ and $\nu =$ 1.25e-5, respectively.  As is the case for the channel flow simulations, a sufficiently fine mesh is employed to fully resolve the mean velocity and pressure fields.

The final flow configuration considered in this paper is turbulent flow over a backward facing step.  DNS data was recently obtained for this flow configuration for an inflow friction Reynolds number of $Re_{\tau} = 395$ and an expansion ratio of 2 \cite{pont2019direct}.  The mean $x$-velocity profile computed from this DNS data is displayed in Figure \ref{Figure10}.  As mentioned previously in this paper, this flow configuration is challenging due to the massive flow separation which results from the stress singularity that occurs at the re-entrant corner, and as illustrated by Figure \ref{Figure1}, even an LEV model employing an ideal viscosity fails to yield accurate predictions of the separation and reattachment points for this flow configuration.  For the simulations reported in this paper, no-slip boundary conditions are applied along the top and bottom walls.  A zero-traction boundary condition is set at the outlet, and fully developed turbulent inflow boundary conditions are set at the inlet by interpolating DNS data.  The density and kinematic viscosity are selected to be $\rho = 1$ and $\nu =$ 7.1e-6, respectively.  As is the case for the channel flow simulations, a sufficiently fine mesh is employed to fully resolve the mean velocity and pressure fields.

\begin{table}[t!]
\caption{Distribution of DNS data for training, validation, and testing.} % title of Table
\centering % used for centering table
\begin{tabular}{|c | c | c | c | c |} % centered columns (4 columns)
\hline 
 &Training &Validation & Testing& Total\\
 \hline
Channel & 96  & 48 & 48 & 192\\
Two-Dimensional Bump & 442,750  & 221,375  & 221,375 & 885,500 \\
Separation Bubble& 656,000  & 328,000 & 328,000 & 1,312,000 \\
Backward Facing Step &228,000 & 114,000 & 114,000 & 456,000 \\
\hline
Total  &1,326,846 &663,423 &663,423 &2,653,692 \\
\hline
\end{tabular}
\label{table:data_distribution} % is used to refer this table in the text
\end{table}

While DNS datasets are often taken to be perfect, this is not always the case due to the large number of time-steps required to obtain sufficiently converged results.  Therefore, we have chosen to smooth the DNS data associated with each of the four selected turbulent flow configurations using a five-point moving average.  We further have interpolated the smoothed DNS data associated with each flow configuration onto a coarser mesh than the DNS mesh.  Finally, all of the interpolated data is then split into three subsets, a training set, a validation set, and a testing set using an approximate 50-25-25 split.  The final number of data points associated with the training, testing, and validation sets for each flow configuration is displayed in Table \ref{table:data_distribution}.

\subsection{Model Training, Validation, and Testing Results}
\label{subsec:trainingresults}
\begin{figure}[t!]
\centering
	\includegraphics[scale=1.0]{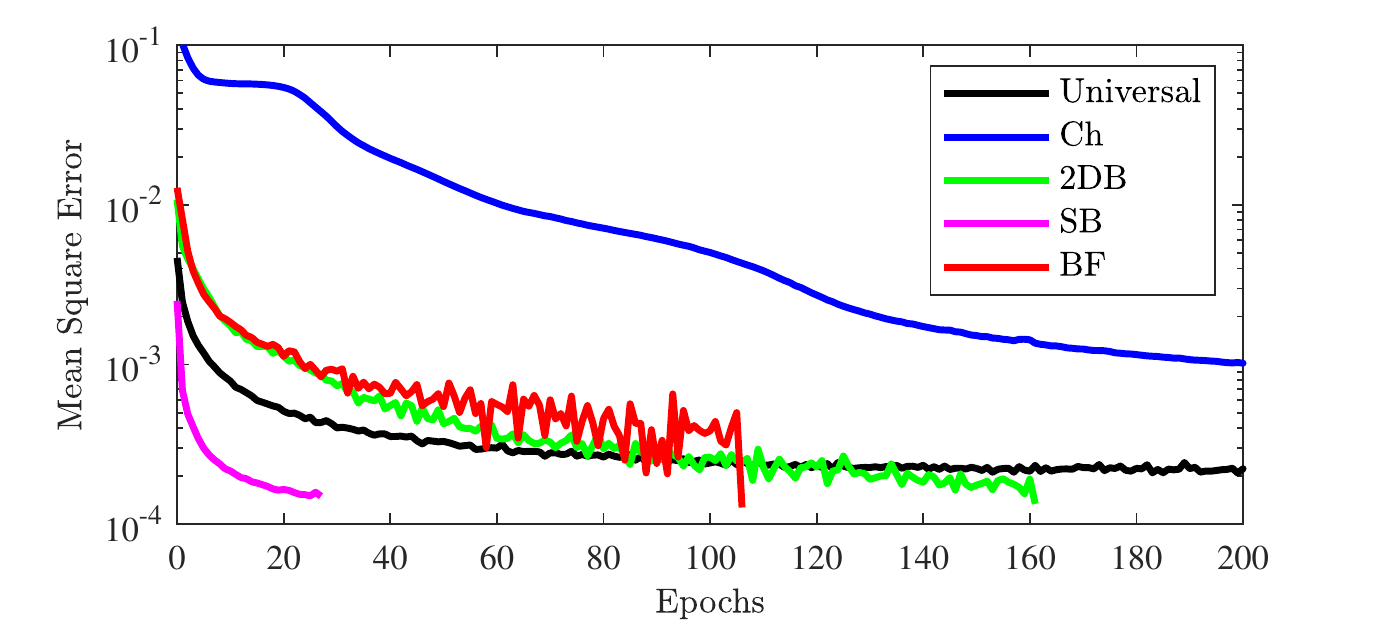}
\caption{Training convergence for the specific (channel (Ch), two-dimensional bump (2DB), separation bubble (SB), and backward facing step (BF)) and universal ANN $S$-frame discrepancy models.}
\label{Figure11}
\end{figure}

We build five ANN $S$-frame discrepancy models in this paper.  Each discrepancy model predicts the discrepancy between the approximate Reynolds stress tensor obtained from the standard $k$-$\varepsilon$ model and the exact Reynolds stress tensor extracted from DNS data.  For the first four discrepancy models, we employ DNS data obtained from one of the four turbulent flow configurations for training, validation, and testing.  We refer to these models as \textit{specific} models, as discussed in Subsection \ref{subsec:learning}.  For the final discrepancy model, we employ DNS data obtained from all four turbulent flow configurations for training, validation, and testing.  We refer to this model as a \textit{universal} model, also as discussed in Subsection \ref{subsec:learning}.  For a particular model case and ANN $S$-frame discrepancy model architecture, we use the procedure discussed in Subsection \ref{subsec:learning} to learn the weights and biases of the corresponding ANN $S$-frame discrepancy model.  We stop the training process when either the training MSE given by \eqref{eq:mse_train} drops below $1.5e-4$ or the number of epochs exceeds 250.  We further adapt the ANN $S$-frame discrepancy model architecture using the procedure discussed in Subsection \ref{subsec:learning}.  Through this adaptation process, we find that a network architecture with 6 hidden layers, 100 neurons per layer, rectified linear unit (ReLU) activation functions for the first 3 hidden layers, and sigmoid activation functions for the last 3 hidden layers yields acceptable results for each specific model case.  Training results are displayed in Figure \ref{Figure11} for the four specific ANN $S$-frame discrepancy models corresponding to this adapted network architecture.  In particular, the MSE is plotted versus number of epochs.  Note that the MSE dips below $1.5e-4$ for each of the specific models except for that associated with channel flow.  Nonetheless, as we shall later see, the specific model returns very accurate results in Reynolds stress propagation tests for the channel test case.   Additionally, we find through adaptation that a network architecture with 6 hidden layers, 200 neurons per layer, rectified linear unit (ReLU) activation functions for the first 3 hidden layers, and sigmoid activation functions for the last 3 hidden layers yields acceptable results for the universal ANN $S$-frame discrepancy model.  Training results are also displayed in Figure \ref{Figure11} for the universal model corresponding to this adapted network architecture.

\afterpage{
\clearpage
\begin{table}[t!]
\caption{Training and testing mean relative error for the channel test case.} % title of Table
\centering % used for centering table
\begin{tabular}{|c | c | c | c | c |} % centered columns (4 columns)
\hline 
 & &$\hat{D}_{11}^S$ &$\hat{D}_{22}^S $& $\hat{D}_{12}^S$\\
\hline
Specific Model & &  & & \\
& Training MRE & 6.0925e-2 & 9.4424e-2 & 5.5328e-2\\
& Testing MRE & 6.1054e-2 & 9.4276e-2 & 5.5513e-2 \\
Universal Model & &  & & \\
& Training MRE & 9.0534e-2 & 9.8931e-2 & 7.5328e-2\\
& Testing MRE & 9.1234e-2 & 9.9761e-2 & 7.6423e-2 \\
\hline
\end{tabular}
\label{table:MRE_channel} % is used to refer this table in the text
\end{table}

\begin{table}[t!]
\caption{Training and testing mean relative error for the two-dimensional bump test case.} % title of Table
\centering % used for centering table
\begin{tabular}{|c | c | c | c | c |} % centered columns (4 columns)
\hline 
 & &$\hat{D}_{11}^S$ &$\hat{D}_{22}^S $& $\hat{D}_{12}^S$\\
\hline
Specific Model & &  & & \\
& Training MRE & 3.4112e-2 & 4.0343e-2 & 1.3489e-2 \\
& Testing MRE & 3.4643e-2 & 4.0857e-2 & 1.3532e-2\\
Universal Model & &  & & \\
& Training MRE & 4.7251e-2  & 5.5437e-2  & 7.9147e-2 \\
& Testing MRE & 4.7474e-2 & 5.5832e-2 & 8.0556e-2 \\
\hline
\end{tabular}
\label{table:MRE_bump} % is used to refer this table in the text
\end{table}

\begin{table}[t!]
\caption{Training and testing mean relative error for the separation bubble test case.} % title of Table
\centering % used for centering table
\begin{tabular}{|c | c | c | c | c |} % centered columns (4 columns)
\hline 
 & &$\hat{D}_{11}^S$ &$\hat{D}_{22}^S $& $\hat{D}_{12}^S$\\
\hline
Specific Model &  & & \\
& Training MRE & 3.7939e-2 & 4.1042e-2 & 1.0628e-1 \\
& Testing MRE & 3.7605e-2 & 4.0889e-2 & 1.0593e-1 \\
Universal Model & &  & & \\
& Training MRE & 5.4577e-2 & 5.3294e-2 & 1.2522e-1\\
& Testing MRE & 5.4722e-2  & 5.3432e-2 & 1.2573e-1  \\
\hline
\end{tabular}
\label{table:MRE_sep_bubble} % is used to refer this table in the text
\end{table}

\begin{table}[t!]
\caption{Training and testing mean relative error for the backward facing step test case.} % title of Table
\centering % used for centering table
\begin{tabular}{|c | c | c | c | c |} % centered columns (4 columns)
\hline 
 & &$\hat{D}_{11}^S$ &$\hat{D}_{22}^S $& $\hat{D}_{12}^S$\\
\hline
Specific Model & &  & & \\
& Training MRE & 5.2619e-2 & 6.0012e-2& 1.6746e-1\\
& Testing MRE & 6.2645e-2 & 7.2145e-2 & 1.6723e-1 \\
Universal Model & &  & & \\
& Training MRE & 1.5211e-1  & 1.6957e-1  & 2.0776e-1 \\
& Testing MRE & 1.5707e-1  & 1.7451e-1  &  2.0890e-1\\
\hline
\end{tabular}
\label{table:MRE_bf_step} % is used to refer this table in the text
\end{table}
\clearpage
}

In Tables \ref{table:MRE_channel}, \ref{table:MRE_bump}, \ref{table:MRE_sep_bubble}, and \ref{table:MRE_bf_step}, we display the MRE in $\hat{D}^S_{11}$, $\hat{D}^S_{22}$, and $\hat{D}^S_{12}$ for both the training and testing datasets for the channel, two-dimensional bump, separation bubble, and backward facing step test cases, respectively.  Moreover, we display the MRE in $\hat{D}^S_{11}$, $\hat{D}^S_{22}$, and $\hat{D}^S_{12}$ for both the specific ANN $S$-frame discrepancy model for the given test case as well as the universal ANN $S$-frame discrepancy model.  Note that, for each case, the training and testing MRE are comparable in size, though as expected, the testing MRE is slightly higher than the training MRE.  The training and testing MRE in $\hat{D}^S_{11}$, $\hat{D}^S_{22}$, and $\hat{D}^S_{12}$ is below 10\% for both the specific and universal ANN $S$-frame discrepancy models for the channel and two-dimensional bump test cases.  The training and testing MRE in $\hat{D}^S_{11}$ and $\hat{D}^S_{22}$ is below 10\% for both the specific and universal models for the separation bubble test case, though the training and validation MRE in $\hat{D}^S_{12}$ is slightly higher.  The training and testing MRE in $\hat{D}^S_{11}$ and $\hat{D}^S_{22}$ is below 10\% for specific model for the backward facing step case, and the training MRE in $\hat{D}^S_{12}$ is slightly higher.  The training and testing MRE in $\hat{D}^S_{11}$, $\hat{D}^S_{22}$, and $\hat{D}^S_{12}$ for the universal model for the backward facing step case is appreciably higher than the specific model.  Nonetheless, as we shall later see, the universal model returns very accurate results in Reynolds stress propagation tests for the backward facing step test case.  While the MRE in $\hat{D}^S_{11}$, $\hat{D}^S_{22}$, and $\hat{D}^S_{12}$ for the validation datasets is not displayed here, the validation MRE is comparable in size to the testing MRE for each case.

\subsection{Reynolds Stress Propagation Results}
\label{subsec:propagationresults}
Even if a particular Reynolds stress closure is highly accurate, it may not yield accurate mean flow field predictions when propagated through the RANS equations \cite{poroseva2016accuracy}.  As such, we employ three types of propagation tests to assess the quality of the five ANN $S$-frame discrepancy models discussed in the previous subsection.
\begin{itemize}[leftmargin=*]
\item \textit{A Priori\footnote{Our notion of \textit{a priori} propagation testing differs from \textit{a priori} tests wherein the Reynolds stress tensor predicted by a given Reynolds stress model is compared with the exact Reynolds stress tensor.} Propagation Test:} We insert the DNS discrepancy tensor,
\begin{gather}
D^{S}_{ij} = D^{S,\text{DNS}}_{ij},
\end{gather}
into the RANS equations and solve for the mean flow field.  The computed mean flow field should match the DNS mean flow field if the simulation mesh is sufficiently fine.
\item \textit{A Posteriori\footnote{In the literature, all Reynolds stress propagation tests are commonly referred to as \textit{a posteriori} tests.  However, it is often unclear whether a given propagation test is of \textit{a priori}, \textit{a posteriori}, or \textit{in situ} type.  Hence, we have elected to use a more precise terminology in this paper.} Propagation Test:} We insert the discrepancy tensor predicted by a given ANN $S$-frame discrepancy model with DNS inputs,
\begin{gather}
D^{S}_{ij} = D^{S,\text{model}}_{ij}\left(\textbf{S}^{\text{DNS}}, \mathbf{\Omega}^{\text{DNS}}, \nabla \overline{p}^{\text{DNS}}, \nabla k^{\text{DNS}}, k^{\text{DNS}}, T^{\text{DNS}}_t, \nu^{\text{DNS}}_t\right),
\end{gather}
into the RANS equations and solve for the mean flow field.  Above, the DNS turbulence time scale is computed as $T^{\text{DNS}}_t = k^{\text{DNS}}/\varepsilon^{\text{DNS}}$ and the DNS turbulence eddy viscosity is computed as $\nu^{\text{DNS}}_t = C_{\mu} \left(k^{\text{DNS}}\right)^2/\varepsilon^{\text{DNS}}$. If the discrepancy model is perfect, the computed mean flow field should match the DNS mean flow field if the simulation mesh is sufficiently fine.  Otherwise, there may be a mismatch.
\item \textit{In Situ Propagation Test:} We insert the discrepancy tensor predicted by a given ANN $S$-frame discrepancy model with simulation mean flow field variable inputs and DNS turbulence variable inputs,
\begin{gather}
D^{S}_{ij} = D^{S,\text{model}}_{ij}\left(\textbf{S}^{\text{simulation}}, \mathbf{\Omega}^{\text{simulation}}, \nabla \overline{p}^{\text{simulation}}, \nabla k^{\text{DNS}}, k^{\text{DNS}}, T^{\text{DNS}}_t, \nu^{\text{DNS}}_t\right),
\end{gather}
into the RANS equations and solve for the mean flow field.  If the discrepancy model is perfect, the computed mean flow field should match the DNS mean flow field if the simulation mesh is sufficiently fine.  Otherwise, there may be a mismatch.
\end{itemize}

\noindent \noindent For each propagation test, we compute the LEV approximation of the anisotropic part of the Reynolds stress tensor using the DNS turbulence eddy viscosity and the simulation mean strain-rate tensor, and we compute the isotropic part of the Reynolds stress tensor using the DNS turbulent kinetic energy.  Unstable results are attained if we instead use the DNS mean strain-rate tensor in computing the LEV approximation of the anisotropic part of the Reynolds stress tensor \cite{wu2019reynolds}.  To assess the quality of the ANN $S$-frame discrepancy models, we compare the results of each propagation test with DNS results.  We also compare with results attained using the Spalart-Allmaras one equation turbulence model \cite{spalart1992one}.

Generally speaking, we expect \textit{a priori} propagation tests to yield the most accurate mean flow field predictions, \textit{a posteriori} propagation tests to yield the next most accurate mean flow field predictions, and \textit{in situ} propagation tests to yield the least accurate mean flow field predictions.  Moreover, we expect \textit{a posteriori}/\textit{in situ} propagation tests employing a specific ANN $S$-frame discrepancy model to yield more accurate mean flow field predictions than \textit{a posteriori}/\textit{in situ} propagation tests employing the universal ANN $S$-frame discrepancy model.  It should also be noted that one more type of discrepancy test can be carried out, namely one where both simulation mean flow field variable inputs and simulation turbulence variable inputs are used with a given ANN $S$-frame discrepancy model.  However, such a test, which we refer to as fully \textit{in situ}, requires the numerical solution of additional model transport equations, one for each turbulence variable.  Unfortunately, such model transport equations also include model closure terms, and as such, it is impossible to determine whether errors associated with a fully \textit{in situ} propagation test are due to inaccuracies in the ANN $S$-frame discrepancy model itself or discrepancies between the model and exact closure terms appearing in the model transport equations.  Consequently, we do not include results obtained from fully \textit{in situ} propagation tests in this paper.

To carry out each of the propagation tests, we solve the RANS equations with the prescribed Reynolds stress model using $P^1$/$P^1$ velocity/pressure finite element approximations, the Streamline Upwind Petrov Galerkin/Pressure Stabilizing Petrov Galerkin (SUPG/PSPG) method, and the open source CFD library \texttt{PHASTA} \cite{whiting2001stabilized,PHASTA}.  Even though each of the selected flow configurations is statistically stationary and two-dimensional, we elect to solve the unsteady RANS equations to steady state for each propagation test.  The eddy viscosity term is handled in an implicit manner in the time advancement of the solution.  This implicit treatment is critical for attaining stable simulation results \cite{wu2019reynolds}.  The discrepancy tensor, on the other hand, is handled in an explicit manner for the \textit{in situ} propagation tests.  Finally, it should be noted that the energetically stable discrepancy model given by \eqref{eq:stable_model} is employed in each of the \textit{in situ} propagation tests as we have observed unstable results are attained otherwise.

The first propagation tests carried out are for the channel test case.  With the channel test case, we are able to determine whether ANN $S$-frame discrepancy models are able to properly predict the law of the wall in the near wall region.   In Figure \ref{Figure12}(a), we plot $u^+$ versus $y^+$ as predicted by the \textit{a priori} propagation test and the \textit{a posteriori} and \textit{in situ} propagation tests using the specific ANN $S$-frame discrepancy model, and we also plot $u^+$ versus $y^+$ as predicted by the DNS and the Spalart-Allmaras turbulence model.  Here, $u^+ = \bar{u}/u_{\tau}$ and $y^+ = yu_{\tau}/\nu$ where $u_{\tau} = \sqrt{\tau_w/\rho}$ is the \textit{friction velocity} and $\tau_w = \left.\left(\rho \nu d\bar{u}/dy\right)\right|_\text{wall}$ is the \textit{wall shear stress}.  From the figure, we see that the results from the \textit{a priori} propagation test and the DNS are indistinguishable.  This indicates that the problem setup is correct and the simulation mesh used in the propagation tests is sufficiently fine as to fully resolve the mean flow field.  We also see that the results from the \textit{a posteriori} and \textit{in situ} tests match those from the DNS, indicating that the specific ANN $S$-frame discrepancy model is able to properly capture the mean flow field in the viscous law region, the buffer layer, and the log-law region.  Finally, the results obtained from the Spalart-Allmaras turbulence model also match those from the DNS.  This is expected as the Spalart-Allmaras turbulence model is known to properly predict the law of the wall in the near wall region \cite{spalart1992one}.  In Figure \ref{Figure12}(b), we plot $u^+$ versus $y^+$ as predicted by the \textit{a posteriori} and \textit{in situ} propagation tests using the universal ANN $S$-frame discrepancy model.  Again, the results from the \textit{a posteriori} and \textit{in situ} tests match those from the DNS, indicating that the universal ANN $S$-frame discrepancy model is also able to properly capture the mean flow field in the viscous law region, the buffer layer, and the log-law region.

\afterpage{
\clearpage
\begin{figure}[t!]
\centering
	\begin{subfigure}[b]{0.48\linewidth}
	        \centering
  		\includegraphics[scale = 1.0]{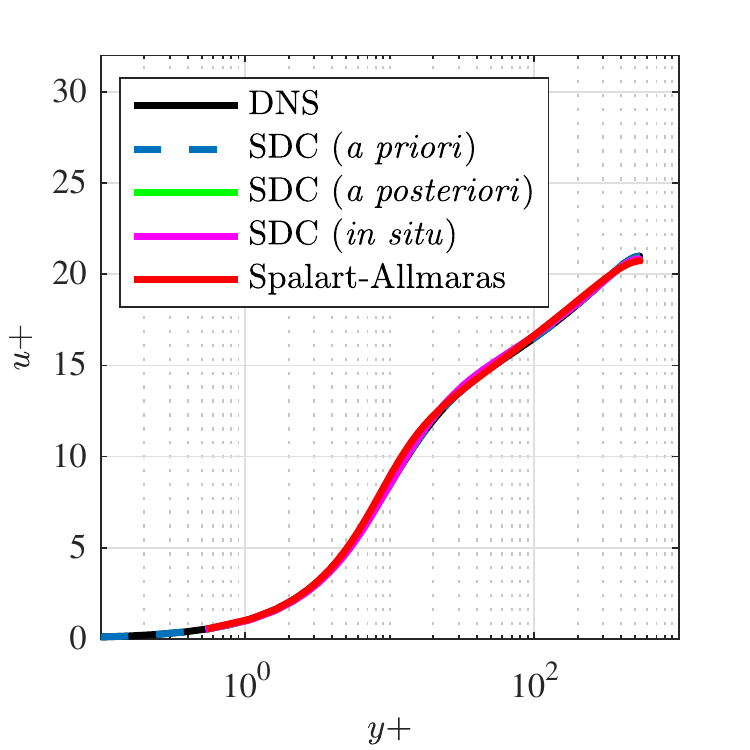}
  		\caption{Specific Model}
	\end{subfigure}
	\begin{subfigure}[b]{0.48\linewidth}
                \centering
  		\includegraphics[scale = 1.0]{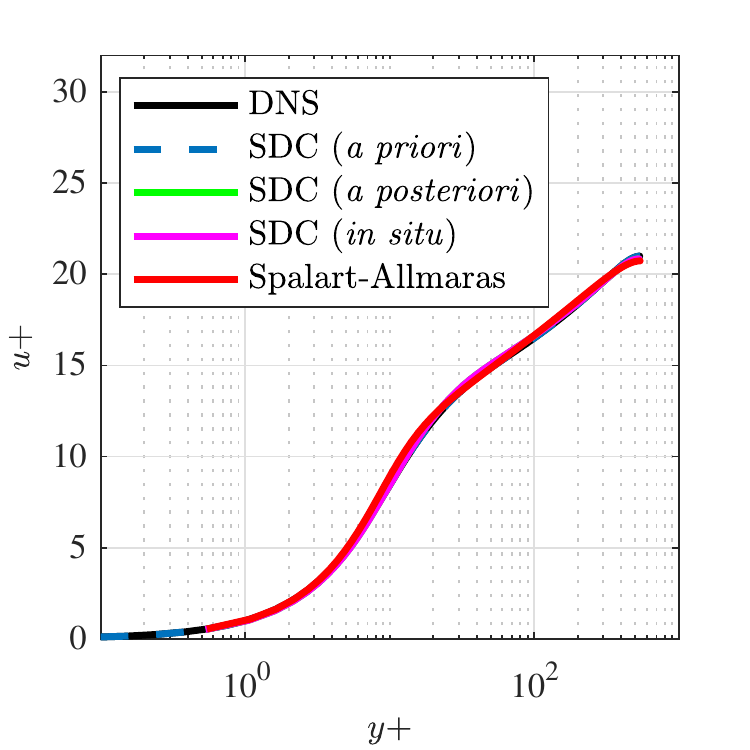}
  		\caption{Universal Model}
	\end{subfigure}
	\caption{Law of the wall as predicted by \textit{a priori}, \textit{a posteriori}, and \textit{in situ} propagation tests using both specific and universal ANN $S$-frame discrepancy (SDC) models.}
	\label{Figure12}
\end{figure}

\begin{figure}[t!]
\centering
	\begin{subfigure}[b]{0.48\linewidth}
		\centering
  		\includegraphics[scale = 1.0]{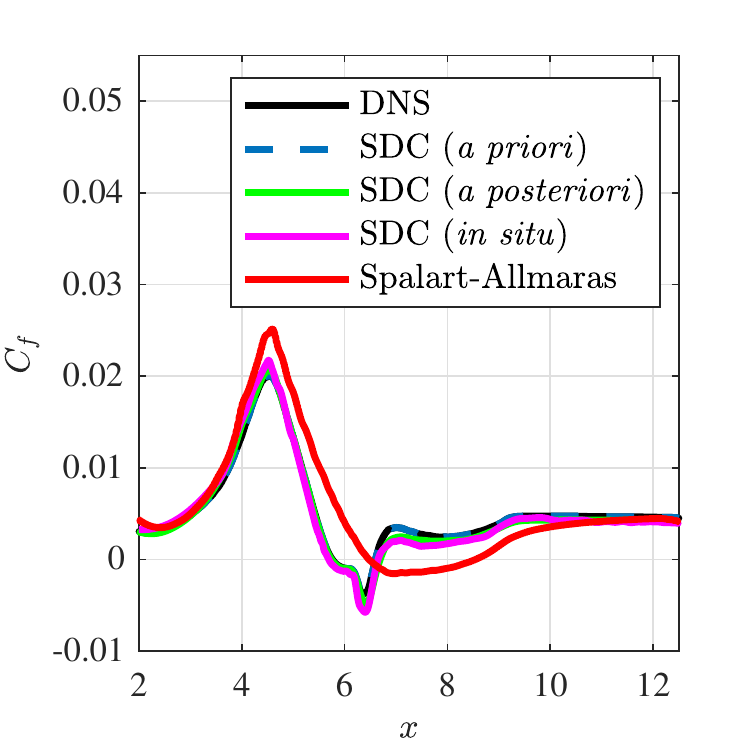}
  		\caption{Specific}
	\end{subfigure}
	\begin{subfigure}[b]{0.48\linewidth}
		\centering
  		\includegraphics[scale = 1.0]{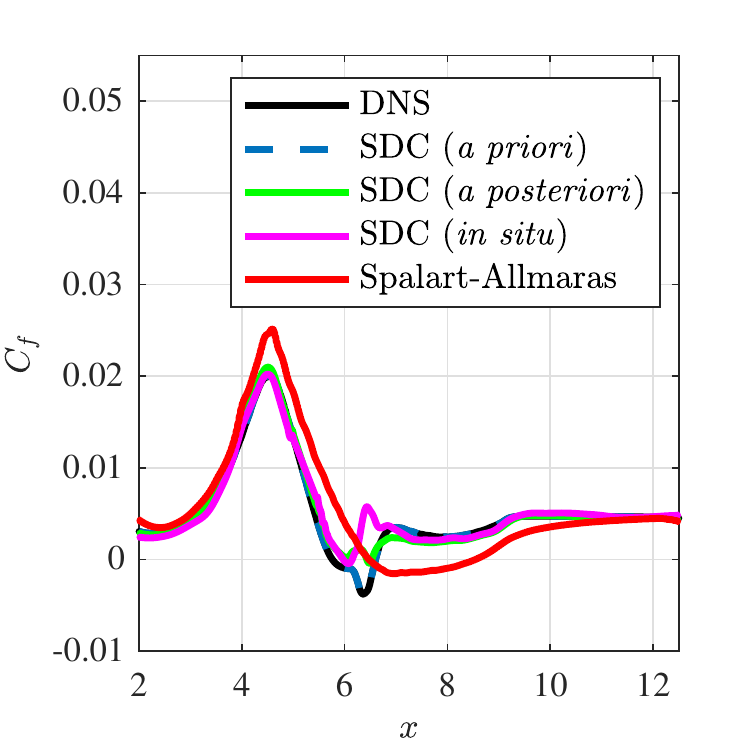}
  		\caption{Universal}
	\end{subfigure}
	\caption{Skin friction coefficient along bottom wall for the two-dimensional bump test case as predicted by \textit{a priori}, \textit{a posteriori}, and \textit{in situ} propagation tests using both specific and universal ANN $S$-frame discrepancy (SDC) models.}
	\label{Figure13}
\end{figure}
\clearpage
}

The next propagation tests carried out are for the two-dimensional bump test case.  With the two-dimensional bump test case, we are able to examine the performance of ANN $S$-frame discrepancy models for a turbulent flow exhibiting shallow flow separation.  In Figure \ref{Figure13}(a), we plot the skin friction coefficient along the bottom wall as predicted by the \textit{a priori} propagation test and the \textit{a posteriori} and \textit{in situ} propagation tests using the specific ANN $S$-frame discrepancy model, and we also plot the skin friction coefficient along the bottom wall as predicted by the DNS and the Spalart-Allmaras turbulence model.  Here, the skin friction coefficient is computed as $C_f = \tau_w / (\frac{1}{2}\rho U^2_\text{max})$ where $U_\text{max}$ is the maximum mean velocity magnitude at the inlet.  From the figure, we see that the results from the \textit{a priori} propagation test and the DNS are indistinguishable, indicating that the problem setup is correct and the simulation mesh used in the propagation tests is sufficiently fine.  We also see that the results from the \textit{a posteriori} and \textit{in situ} tests match those from the DNS, suggesting that the specific ANN $S$-frame discrepancy model is able to properly capture the mean flow field in the separation region.  In particular, the \textit{a posteriori} and \textit{in situ} tests yield accurate predictions of the separation and reattachment locations (defined here as the locations where the skin friction coefficient changes sign from positive to negative and negative to positive, respectively).  The Spalart-Allmaras turbulence model, on the other hand, fails to accurately predict the skin friction coefficient along the bottom wall as well as the separation and reattachment locations.  In Figure \ref{Figure13}(b), we plot the skin friction coefficient along the bottom wall as predicted by the \textit{a posteriori} and \textit{in situ} propagation tests using the universal ANN $S$-frame discrepancy model.  The \textit{a posteriori} test using the universal ANN $S$-frame discrepancy model yields accurate predictions the skin friction coefficient before and after the separation region, and it accurately predicts the separation and reattachment locations.  However, it fails to accurately predict the skin friction coefficient within the separation region.  The results of the \textit{in situ} test are marginally less accurate than the results of the \textit{a posteriori} test, though they are considerably more accurate than the results obtained using the Spalart-Allmaras turbulence model outside the separation region.

\afterpage{
\clearpage
\begin{figure}[t!]
\centering
	\begin{subfigure}[b]{0.48\linewidth}
	        \centering
  		\includegraphics[scale= 1.0]{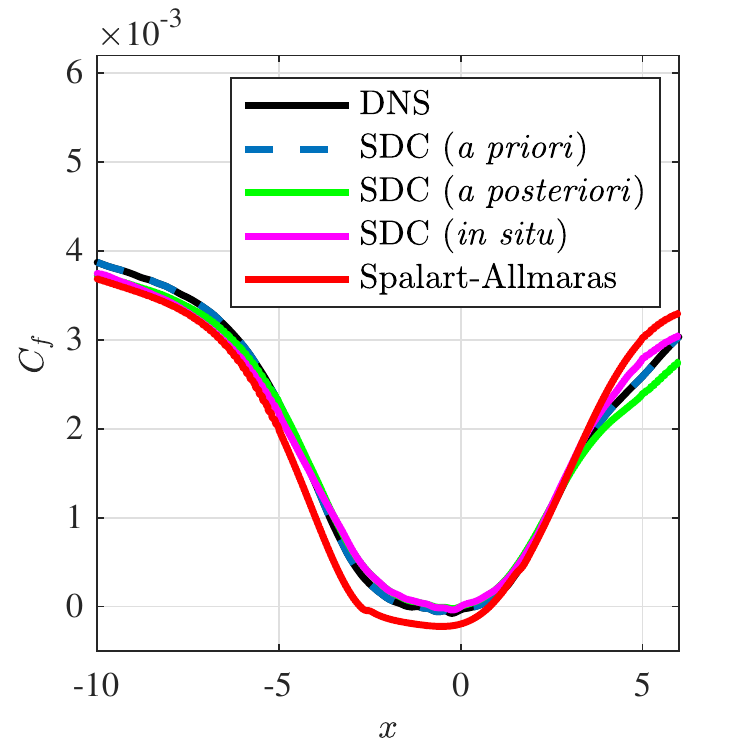}
  		\caption{Specific}
	\end{subfigure}
	\begin{subfigure}[b]{0.48\linewidth}
                \centering
  		\includegraphics[scale= 1.0]{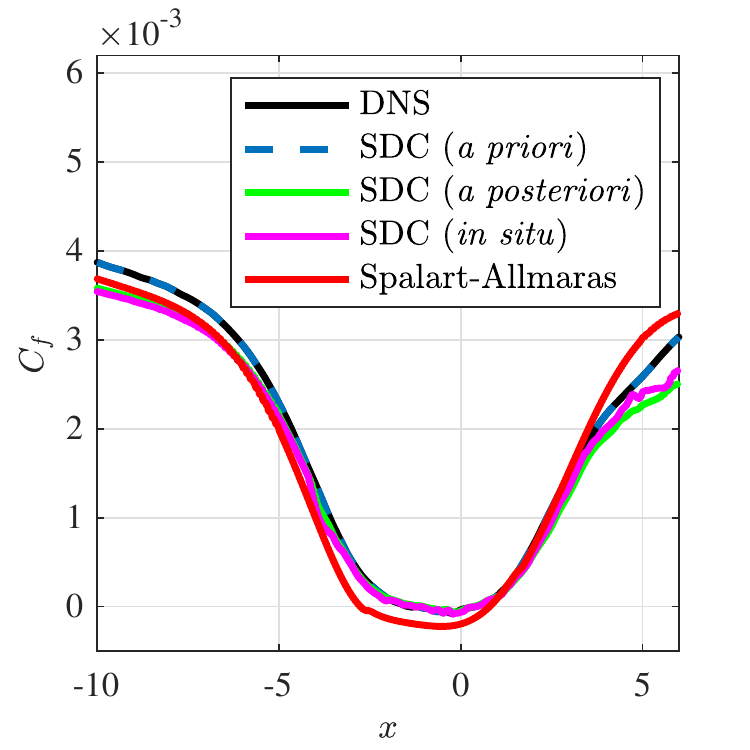}
  		\caption{Universal}
	\end{subfigure}
	\caption{Skin friction coefficient along bottom wall for the separation bubble test case as predicted by \textit{a priori}, \textit{a posteriori}, and \textit{in situ} propagation tests using both specific and universal ANN $S$-frame discrepancy (SDC) models.}
	\label{Figure14}
\end{figure}

\begin{figure}[t!]
\centering
	\begin{subfigure}[b]{0.48\linewidth}
	        \centering
  		\includegraphics[scale= 1.0]{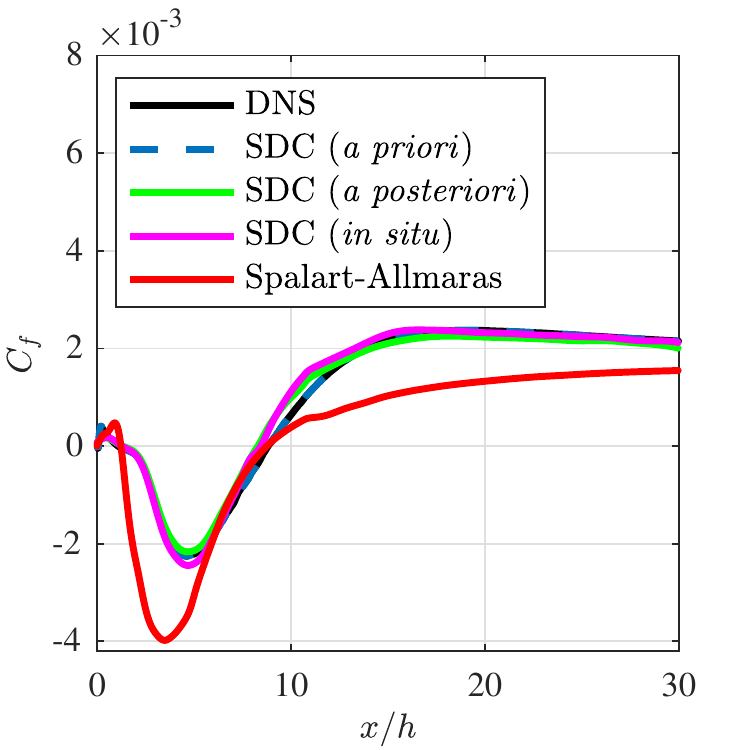}
  		\caption{Specific}
	\end{subfigure}
	\begin{subfigure}[b]{0.48\linewidth}
                \centering
  		\includegraphics[scale= 1.0]{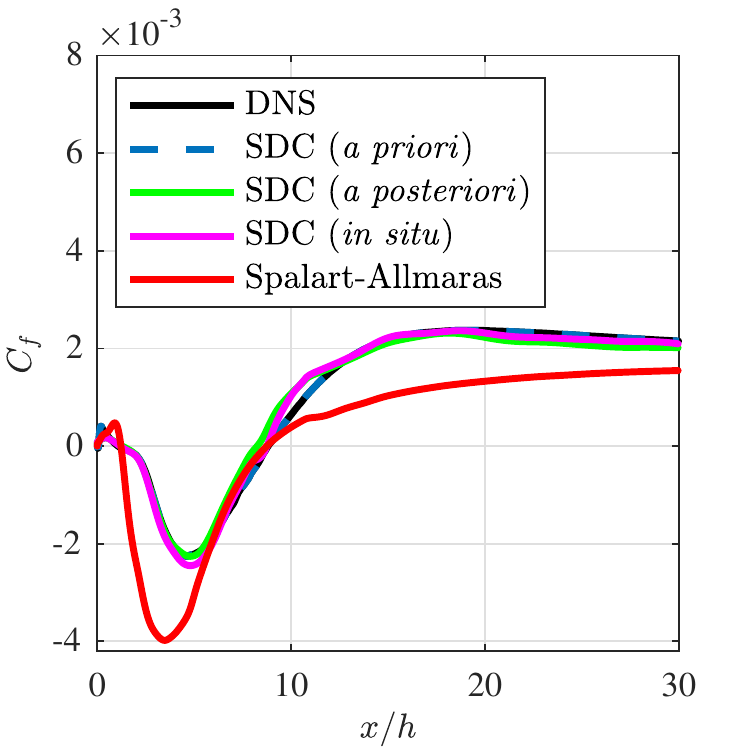}
  		\caption{Universal}
	\end{subfigure}
	\caption{Skin friction coefficient along bottom wall for the backward facing step test case as predicted by \textit{a priori}, \textit{a posteriori}, and \textit{in situ} propagation tests using both specific and universal ANN $S$-frame discrepancy (SDC) models.}
	\label{Figure15}
\end{figure}
\clearpage
}

The next propagation tests carried out are for the separation bubble test case.  This test case is interesting as the separation bubble is not a result of an adverse pressure gradient due to curvature but rather suction and blowing along the upper wall.  In fact, this test case was carefully designed to induce minimal flow separation on the bottom wall, and as a consequence, some state-of-the-art RANS models fail to predict separation at all \cite{coleman2018numerical}.  In Figure \ref{Figure14}(a), we plot the skin friction coefficient along the bottom wall as predicted by the \textit{a priori}, \textit{a posteriori}, and \textit{in situ} propagation tests using the specific ANN $S$-frame discrepancy model, and we also plot the skin friction coefficient along the bottom wall as predicted by the DNS and the Spalart-Allmaras turbulence model.  Here, the skin friction coefficient is computed as $C_f = \tau_w / (\frac{1}{2}\rho U^2_\infty)$ where $U_\infty$ is the free-stream mean velocity magnitude associated with the zero pressure gradient boundary layer solution applied at the inlet.  The results from the \textit{a priori} propagation test and the DNS are again indistinguishable, indicating that the problem setup is correct and the simulation mesh used in the propagation tests is sufficiently fine.  We also see that the results from the \textit{a posteriori} and \textit{in situ} tests closely match those from the DNS, except in a small region near the inlet and outlet.  In particular, the \textit{a posteriori} and \textit{in situ} test results and the DNS results are visibly indistinguishable near the location where the skin friction coefficient is approximately zero.   We hypothesize that the discrepancy between the \textit{a posteriori} and \textit{in situ} test data and the DNS data near the inlet and outlet might be eliminated with an alternate selection of boundary condition along the upper wall.  The Spalart-Allmaras turbulence model fails to accurately predict the skin friction coefficient throughout the channel, and in particular, it predicts the presence of a secondary long, shallow separation bubble along the bottom wall, which is non-physical.  In Figure \ref{Figure14}(b), we plot the skin friction coefficient along the bottom wall as predicted by the \textit{a posteriori} and \textit{in situ} propagation tests using the universal ANN $S$-frame discrepancy model.  From the figure, it is seen that the \textit{a posteriori} and \textit{in situ} tests using the universal ANN $S$-frame discrepancy model yield comparable predictions to the {a posteriori} and \textit{in situ} tests using the specific ANN $S$-frame discrepancy model.

The final propagation tests carried out are for the backward facing step test case.  This test case is challenging for RANS models due to the massive flow separation that occurs downstream of the re-entrant corner.  In Figure \ref{Figure15}(a), we plot the skin friction coefficient along the bottom wall as predicted by the \textit{a priori} propagation test and the \textit{a posteriori} and \textit{in situ} propagation tests using the specific ANN $S$-frame discrepancy model, and we also plot the skin friction coefficient along the bottom wall as predicted by the DNS and the Spalart-Allmaras turbulence model.  Here, the skin friction coefficient is computed as $C_f = \tau_w / (\frac{1}{2}\rho U^2_\text{max})$ where $U_\text{max}$ is the maximum velocity at the inlet, as was the case for the two-dimensional bump test case.  Like the first three sets of propagation tests, the results from the \textit{a priori} propagation test and DNS are indistinguishable, indicating that the simulation mesh used in the propagation tests is sufficiently fine.  We also see that the results from the \textit{a posteriori} and \textit{in situ} tests using the specific ANN $S$-frame discrepancy model nearly match those from DNS.  In particular, the \textit{a posteriori} and \textit{in situ} tests yield accurate predictions of the reattachment location.  The Spalart-Allmaras turbulence model also yields accurate predictions of the reattachment location, but it fails to accurately predict the skin friction coefficient at all other locations along the bottom wall.  As we will see in the next section, the Spalart-Allmaras turbulence model also produces an inaccurate mean velocity field profile throughout the simulation domain.  In Figure \ref{Figure15}(b), we plot skin friction coefficient along the bottom wall as predicted by the \textit{a posteriori} and \textit{in situ} propagation tests using the universal ANN $S$-frame discrepancy model.  The results of both the \textit{a posteriori} and \textit{in situ} tests using the universal ANN $S$-frame discrepancy model nearly match those from DNS, and both tests yield accurate predictions of the reattachment location.

\section{Sensitivity and Dimensional Reduction}
\label{sec:sensitivity}
One of the advantages of having an ANN $S$-frame discrepancy model is the ability to compute sensitivities with respect to model inputs.  These sensitivities may be employed to reduce the dimension of the input feature space and consequently reduce computational cost associated with the discrepancy model.  Sensitivities may alternatively guide engineers in the design of explicit analytical $S$-frame discrepancy models \cite{schmelzer2020discovery}.  Such discrepancy models would be much simpler to analyze than ANN $S$-frame discrepancy models, and they also would be much easier to implement in state-of-the-art CFD codes.

There are many different measures of sensitivity that one can employ (see, e.g., \cite{saltelli2004sensitivity}).  We elect here to use \textit{total sensitivity indices}, first introduced by Sobol in \cite{sobol92sensitivity}, since they measure sensitivity across the entire global input space, they can deal with nonlinear responses, and they can account for interactions between input variables.  Total sensitivity indices can also be computed in an efficient manner as compared with other commonly employed sensitivity measures.

For a general square-integrable function $g:[-1,1]^m \rightarrow \mathbb{R}$, let the expectation of $g$ be defined as
\begin{equation}
\text{E}[g] = 2^{-m} \int_{[-1,1]^m} g(\textbf{x}) d\textbf{x},
\end{equation}
and let the variance of $g$ be defined as
\begin{equation}
\text{Var}[g] = \text{E}\left[ \left( g(\textbf{x}) - \text{E}(g) \right)^2 \right].
\end{equation}
Now, let $f: [-1,1]^m \rightarrow \mathbb{R}$ be a particular square-integrable function.  For subsets $u \subseteq \{1, \ldots, m\}$, let $|u|$ be the cardinality of $u$, $u_c$ be the complement of $u$ in $\{1, \ldots, m\}$, and $\textbf{x}^u$ be the $|u|$-tuple of components $x_j$ for $j \in u$.  Define the functions $f_u: [-1,1]^m \rightarrow \mathbb{R}$ recursively as
\begin{equation}
f_u(\textbf{x}) = 2^{-|u_c|} \int_{[-1,1]^{|u_c|}} f(\textbf{x}) d\textbf{x}^{u_c} - \sum_{v \subsetneq u} f_v(\textbf{x}),
\end{equation}
where
\begin{equation}
f_\emptyset(\textbf{x}) = \text{E}[f].
\end{equation}
Note that $f_u(\textbf{x})$ only depends on $\textbf{x}$ through $\textbf{x}^u$.  Finally, define the set $\mathcal{S}_i$ as the set of subsets of $\{1, \ldots, m\}$ containing the index $i$.  The total sensitivity index $S^\text{tot}_i$ for the $i^\text{th}$ input variable $x_i$ is then defined as
\begin{equation}
S^\text{tot}_i = \frac{\sum_{u \subseteq \mathcal{S}_i} \text{Var}[f_u]}{\text{Var}[f]}.
\end{equation}
Briefly speaking, the total sensitivity index $S^\text{tot}_i$ measures the contribution of the $i^\text{th}$ input variable $x_i$ to the overall variance of $f$, including all variance caused by the interaction of $x_i$ with other input variables.  By construction, $0 \leq S^\text{tot}_i \leq 1$, and the larger $S^\text{tot}_i$ is, the more $x_i$ contributes to the overall variance of $f$.  When $f$ is smooth and can be evaluated quickly and $m$ is relatively small, one can efficiently and accurately compute $S^\text{tot}_i$ using numerical quadrature.  Otherwise, one can compute $S^\text{tot}_i$ using the Monte Carlo method.  We elect to use the Monte Carlo method to compute total sensitivity indices for ANN $S$-frame discrepancy models as these models have a large number of inputs.  In particular, we use the Monte Carlo method introduced by Jansen in \cite{jansen1999analysis}.

\afterpage{
\clearpage
\begin{table}[t!]
\caption{Total sensitivity indices for the specific ANN $S$-frame discrepancy models corresponding to the two-dimensional bump (2DB), separation bubble (SB), and backward facing step (BF) test cases.}
\centering % used for centering table
\begin{tabular}{|c | c | c | c |c | c |c | c | c | c|} % centered columns (4 columns)
\hline 
\cellcolor{blue!25}2DB &\multicolumn{1}{c}{} & \multicolumn{1}{c}{} &\multicolumn{1}{c}{}  & & \cellcolor{blue!25}SB &\multicolumn{1}{c}{} &\multicolumn{1}{c}{} &\multicolumn{1}{c}{} & \\% inserts table
\hline 
\hline 
Input & $\hat{D}_{11}^S$ & $\hat{D}_{22}^S$ & $\hat{D}_{12}^S$ & Average & Input & $\hat{D}_{11}^S$ & $\hat{D}_{22}^S$ & $\hat{D}_{12}^S$ & Average \\% inserts table
\hline 
    1  &  0.3019 &   0.3112 &   0.1091 &\cellcolor{green!25}   0.2407 &  1  &   0.3577 &   0.2770 &   0.0614 & \cellcolor{green!25}  0.2320 \\
    2   & 0.0076 &   0.0032 &   0.0022 &   0.0043 &                                2 &    0.0034 &    0.0012 &    0.0094 &    0.0046\\
    3   & 0.0796  &  0.0630 &   0.0999 &\cellcolor{green!25}   0.0809 &  3 &    0.0532 &    0.0560 &    0.0589 & \cellcolor{green!25}   0.0560\\
    4  &  0.0826 &   0.0803 &   0.0792 &\cellcolor{green!25}   0.0807 &    4 &    0.0488 &    0.0658 &    0.1528 &\cellcolor{green!25}    0.0891\\
    5  &  0.0599 &   0.0862 &   0.0828 &\cellcolor{green!25}   0.0763 &    5 &    0.0689 &    0.0763 &    0.1402 &\cellcolor{green!25}    0.0951\\
    6  &  0.0667 &   0.0685 &   0.0856 &\cellcolor{green!25}   0.0736 &    6 &    0.0266 &    0.0459 &    0.0284 &    0.0336\\
    7  &  0.0898 &   0.0908 &   0.0916 &\cellcolor{green!25}   0.0907 &    7 &    0.0687 &    0.1048 &    0.1061 & \cellcolor{green!25}   0.0932\\
    8  &  0.0341 &   0.0293 &   0.0766 & \cellcolor{green!25}  0.0467 &   8 &    0.0100 &    0.0145 &    0.0185 &    0.0143\\
    9   & 0.0181 &   0.0200 &   0.0272 &   0.0218 &                                  9 &    0.0448 &    0.0597 &    0.0338 &\cellcolor{green!25}    0.0461\\
   10  &  0.0427 &   0.0477 &   0.0718 & \cellcolor{green!25}  0.0541 &   10 &    0.0771 &    0.0654 &    0.0478 & \cellcolor{green!25}   0.0634\\
   11  &  0.0185 &   0.0197 &   0.0574  &  0.0318 &                                  11 &    0.0305 &    0.0537 &    0.1173 &  \cellcolor{green!25}  0.0672\\
   12  &  0.0118 &   0.0095 &   0.0331 &   0.0181 &                                 12 &    0.0400 &    0.0304 &    0.0270 &    0.0324\\
   13  &  0.0157 &   0.0142 &   0.0183 &   0.0161 &                                13 &    0.0573 &    0.0499 &    0.0488 & \cellcolor{green!25}   0.0520\\
   14  &  0.1193  &  0.0938 &   0.0728 &\cellcolor{green!25}   0.0953 &   14 &    0.0404 &    0.0356 &    0.0487 &    0.0416\\
   15  &  0.0593 &   0.0657 &   0.0946 &\cellcolor{green!25}   0.0732 &   15 &    0.0760 &    0.0649 &    0.1104 & \cellcolor{green!25}   0.0838\\

 \hline
 \hline
\cellcolor{blue!25}BF &\multicolumn{1}{c}{} & \multicolumn{1}{c}{} &\multicolumn{1}{c}{}  &\multicolumn{1}{c}{} &\multicolumn{1}{c}{}  &\multicolumn{1}{c}{} &\multicolumn{1}{c}{} & \multicolumn{1}{c}{}&\\
\hline
\hline
Feature & $\hat{D}_{11}^S$ & $\hat{D}_{22}^S$ & $\hat{D}_{12}^S$ & Average &\multicolumn{1}{c}{} &\multicolumn{1}{c}{} & \multicolumn{1}{c}{}& & Total Average  \\
\hline
    1 &   0.1475 &    0.1540 &    0.1136 &\cellcolor{green!25}    0.1384  &\multicolumn{1}{c}{} &\multicolumn{1}{c}{} & \multicolumn{1}{c}{}& &  \cellcolor{green!25}0.2037\\
    2  &  0.0035 &    0.0038 &    0.0062 &    0.0045 &\multicolumn{1}{c}{} &\multicolumn{1}{c}{} & \multicolumn{1}{c}{}& &    0.0015  \\
    3  &  0.0601 &    0.0611 &    0.0941 & \cellcolor{green!25}   0.0718  &\multicolumn{1}{c}{} &\multicolumn{1}{c}{} & \multicolumn{1}{c}{}& & \cellcolor{green!25}   0.0696  \\
    4 &    0.1109 &    0.1400 &    0.0839 &\cellcolor{green!25}    0.1116  &\multicolumn{1}{c}{} &\multicolumn{1}{c}{} & \multicolumn{1}{c}{}& & \cellcolor{green!25}   0.0938  \\
    5 &    0.0603 &    0.0961 &    0.0730 & \cellcolor{green!25}   0.0764  &\multicolumn{1}{c}{} &\multicolumn{1}{c}{} & \multicolumn{1}{c}{}& & \cellcolor{green!25}    0.0826 \\
    6 &    0.0468 &    0.0591 &    0.0631 &\cellcolor{green!25}    0.0563  &\multicolumn{1}{c}{} &\multicolumn{1}{c}{} & \multicolumn{1}{c}{}& &\cellcolor{green!25}     0.0545 \\
    7 &    0.0866 &    0.0785 &    0.1098 & \cellcolor{green!25}   0.0916 &\multicolumn{1}{c}{} &\multicolumn{1}{c}{} & \multicolumn{1}{c}{}& & \cellcolor{green!25}    0.0919 \\
    8 &    0.0277 &    0.0438 &    0.0345 &    0.0353  &\multicolumn{1}{c}{} &\multicolumn{1}{c}{} & \multicolumn{1}{c}{}& &     0.0321 \\
    9  &   0.0278 &    0.0126 &    0.0316 &    0.0240  &\multicolumn{1}{c}{} &\multicolumn{1}{c}{} & \multicolumn{1}{c}{}& &     0.0306 \\
   10 &    0.0387 &    0.0434 &    0.0553 & \cellcolor{green!25}   0.0458  &\multicolumn{1}{c}{} &\multicolumn{1}{c}{} & \multicolumn{1}{c}{}& & \cellcolor{green!25}     0.0544\\
   11 &    0.0469 &    0.0487 &    0.0538 &\cellcolor{green!25}    0.0498  &\multicolumn{1}{c}{} &\multicolumn{1}{c}{} & \multicolumn{1}{c}{}& &\cellcolor{green!25}      0.0496\\
   12 &    0.0302 &    0.0217 &    0.0515 &    0.0344  &\multicolumn{1}{c}{} &\multicolumn{1}{c}{} & \multicolumn{1}{c}{}& &     0.0283 \\
   13 &    0.0396 &    0.0325 &    0.0395 &    0.0372  &\multicolumn{1}{c}{} &\multicolumn{1}{c}{} & \multicolumn{1}{c}{}& &     0.0351 \\
   14 &    0.1873 &    0.0617 &    0.1020 &\cellcolor{green!25}    0.1170  &\multicolumn{1}{c}{} &\multicolumn{1}{c}{} & \multicolumn{1}{c}{}& &\cellcolor{green!25}      0.0846\\
   15 &   0.0861 &    0.1429 &    0.0880 &\cellcolor{green!25}    0.1057 &\multicolumn{1}{c}{} &\multicolumn{1}{c}{} & \multicolumn{1}{c}{}& &\cellcolor{green!25}      0.0876\\
   \hline
\end{tabular}
\label{table:sobol_rankings} % is used to refer this table in the text
\end{table}
\clearpage
}

In Table \ref{table:sobol_rankings}, we report the total sensitivity indices associated with the specific ANN $S$-frame discrepancy models constructed in the previous section for the two-dimensional bump, separation bubble, and backward facing step test cases.  For each test case, we report total sensitivity indices for the ANN models for $\hat{D}^{S}_{11}$, $\hat{D}^{S}_{22}$, and $\hat{D}^{S}_{12}$, and we calculate the average sensitivity index across these three ANN models for each input variable.  Moreover, we calculate the average sensitivity index across all three test cases.  Finally, we highlight the inputs corresponding to the ten highest average sensitivity indices for each test case and across all test cases in green.  As each test case is statistically two-dimensional, each ANN model has only 15 model inputs, those highlighted in blue in Table \ref{table:invariant_inputs}.  Note that the first model input (i.e., $\hat{q}_1 = \hat{S}_{ij}\hat{S}_{ij} = \hat{\textbf{S}}:\hat{\textbf{S}}$) has by far the highest total sensitivity index for the ANN models for $\hat{D}^{S}_{11}$ and $\hat{D}^{S}_{22}$ for each test case.  This indicates that mean strain-rate magnitude is a critical ingredient in the design of an accurate explicit analytical model for $\hat{D}^{S}_{11}$ and $\hat{D}^{S}_{22}$.  This is not too surprising, as mean strain-rate magnitude is often a key ingredient in Reynolds stress closures.  However, a number of other model inputs have high total sensitivity indices for the ANN models for $\hat{D}^{S}_{12}$.  In particular, the fourth (i.e., $\hat{q}_5 = \hat{A}^k_{ij} \hat{A}^k_{ji} = \widehat{\nabla k} \cdot \widehat{\nabla k}$), fifth (i.e., $\hat{q}_9 = \hat{A}^p_{ij} \hat{A}^p_{jk} \hat{S}_{ki} = \widehat{\nabla \bar{p}} \cdot \hat{\textbf{S}} \cdot \widehat{\nabla \bar{p}}$), and seventh (i.e., $\hat{q}_{16} = \hat{A}^p_{ij} \hat{A}^k_{jk} = \widehat{\nabla \bar{p}} \cdot \widehat{\nabla k}$) model inputs have high total sensitivity indices for each test case, nearly as high or even exceeding the total sensitivity index associated with the first model input.  Consequently, these model inputs should be included in the design of an accurate explicit analytical model for $\hat{D}^{S}_{12}$.

\begin{table}[t!]
\caption{Inputs corresponding to the ten highest average total sensitivity indices.} % title of Table
\centering % used for centering table
\begin{tabular}{|c | c | c | c | c | c |} % centered columns (4 columns)
\hline &&&&& \\ [-1em] %inserts double horizontal lines
Input  & Relation & Input  & Relation & Input  & Relation  \\ [0.5ex] % inserts table
%heading
\hline &&&&& \\ [-1em] % inserts single horizontal line
$\hat{q}_{1}$ & $\sh_{ij} \sh_{ji} $ & $\hat{q}_{9}$ &  $\ph_{ij}\ph_{jk} \sh_{ki}$ & $\hat{q}_{34}$  & $\ph_{ij}\kh_{jk} \sh_{ki}$  \\ % inserting body of the table
$\hat{q}_4$ & $\ph_{ij}\ph_{ji}$ & $\hat{q}_{12}$ &  $\kh_{ij}\kh_{jk} \sh_{ki}$ & $\hat{q}_{42}$  &  $\oh_{ij} \ph_{jk} \kh_{ki}$ \\ % inserting body of the table
$\hat{q}_{5}$ &  $\kh_{ij}\kh_{ji}$ & $\hat{q}_{16}$  &  $\ph_{ij}\kh_{ji}$ & $\hat{q}_{49}$, $\hat{q}_{50}$ & $\hat{T}_t$, $\hat{\nu}_t$ \\ [1ex] % inserting body of the table
\hline %inserts single linet
\end{tabular}
\label{table:sobol_highest}
\end{table}

\begin{table}[t!]
\caption{Training and testing mean relative error for the backward facing step case and the universal ANN $S$-frame discrepancy model with only 10 model inputs.} % title of Table
\centering % used for centering table
\begin{tabular}{|c | c | c | c |} % centered columns (4 columns)
\hline 
 &$\hat{D}_{11}^S$ &$\hat{D}_{22}^S $& $\hat{D}_{12}^S$\\
\hline
Training Error & 1.2010e-1  & 1.3125e-1 & 1.1169e-1 \\
Testing Error & 1.2505e-1 & 1.3712e-1  & 1.1998e-1 \\
\hline
\end{tabular}
\label{table:MRE_dim_red} % is used to refer this table in the text
\end{table}

\begin{figure}[t!]
\centering
        \centering
  		\includegraphics[scale= 1.0]{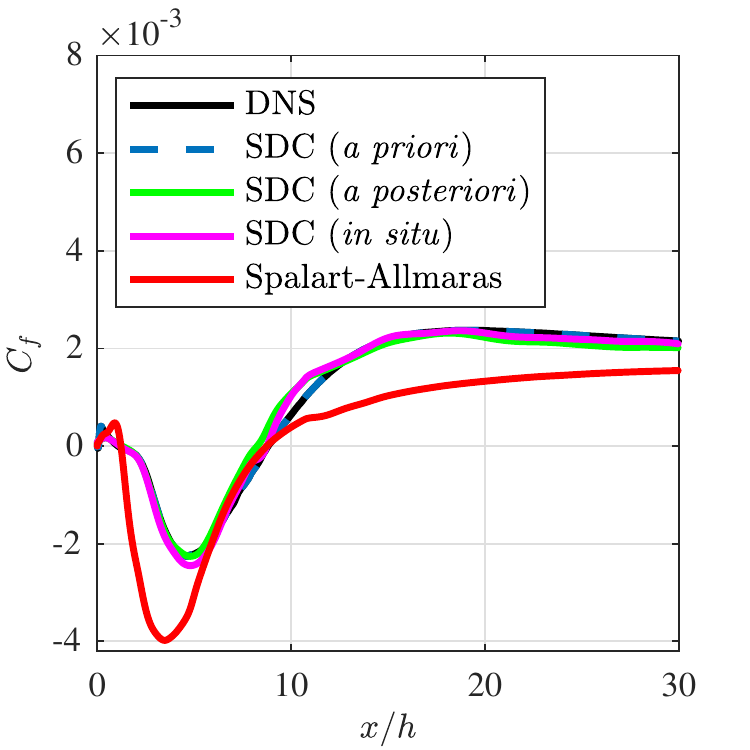}
	\caption{Skin friction coefficient along bottom wall for the backward facing step test case as predicted by \textit{a priori}, \textit{a posteriori}, and \textit{in situ} propagation tests using the universal ANN $S$-frame discrepancy (SDC) model with only 10 model inputs.}
	\label{Figure16}
\end{figure}

\begin{figure}[t!]
  \centering
  \includegraphics[scale=0.77]{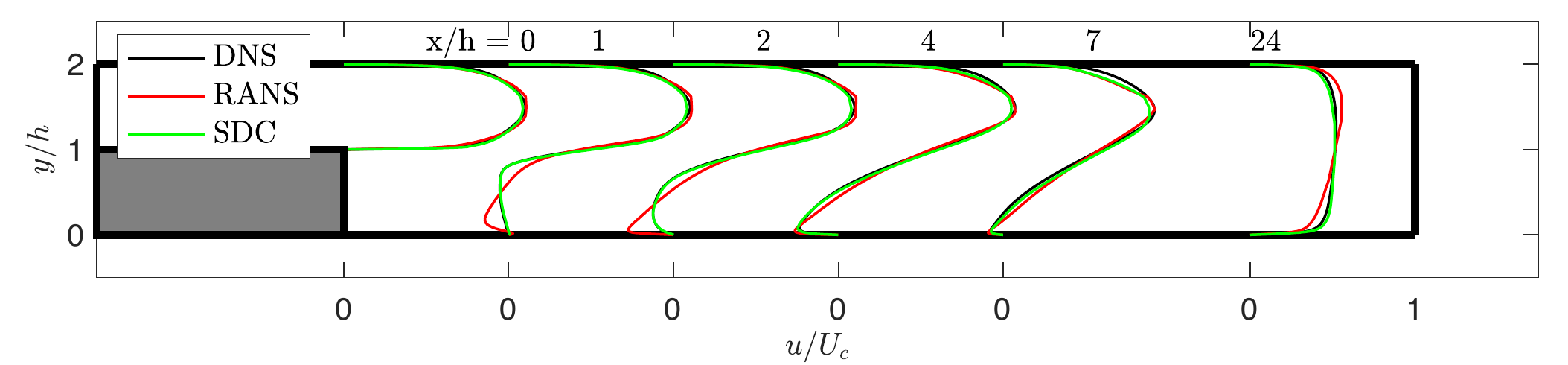}
\caption{Mean $x$-velocity profiles for the backward facing step case as predicted using DNS, the Spalart-Allmaras (RANS) turbulence model, and an \textit{in situ} propagation test using the universal ANN $S$-frame discrepancy (SDC) model with only 10 model inputs.}
\label{Figure17}
\end{figure}

We now construct a universal ANN $S$-frame discrepancy model using only the 10 model inputs with highest average total sensitivity index.  These inputs are displayed in Table \ref{table:sobol_highest}.  We employ a network architecture with 6 hidden layers, 200 neurons per layer, ReLU activation functions for the first 3 hidden layers, and sigmoid activation functions for the last 3 hidden layers.  We train the discrepancy model using less training data than in Section \ref{sec:numerical}  in order to equally weight the two-dimensional bump, separation bubble, and backward facing step test cases.  In Table \ref{table:MRE_dim_red}, we display the MRE in $\hat{D}^S_{11}$, $\hat{D}^S_{22}$, and $\hat{D}^S_{12}$ for both the training and testing datasets for the backward facing step test case and the universal ANN $S$-frame discrepancy model with only 10 model inputs.  Note that the MRE is comparable in size to that associated with the universal ANN $S$-frame discrepancy model with all 15 model inputs for each of $\hat{D}^S_{11}$, $\hat{D}^S_{22}$, and $\hat{D}^S_{12}$.  This suggests that model fidelity has not been sacrificed with dimensional reduction of the input space.  In fact, the accuracy is even better with the universal ANN $S$-frame discrepancy model with 10 model inputs in some cases.  This is due to the fact that the discrepancy model was trained using an equal number of samples from the two-dimensional bump, separation bubble, and backward facing step datasets.  In Figure \ref{Figure16}, we plot the skin friction coefficient along the bottom wall as predicted by an \textit{a priori} propagation test and \textit{a posteriori} and \textit{in situ} propagation tests using the universal ANN $S$-frame discrepancy model with only 10 model inputs, and we also plot the skin friction coefficient along the bottom wall as predicted by the DNS and the Spalart-Allmaras turbulence model.  The skin friction coefficient is predicted accurately using both the \textit{a posteriori} and \textit{in situ} propagation tests, despite dimensional reduction of the input space.  Finally, in Figure \ref{Figure17}, mean $x$-velocity profiles at different locations downstream from the step as predicted using DNS, the \textit{in situ} propagation test, and the Spalart-Allmaras turbulence model are displayed.  The DNS and \textit{in situ} propagation test mean $x$-velocity profiles are almost indistinguishable, while the Spalart-Allmaras turbulence model mean $x$-velocity profile is much less accurate.

\section{Conclusions}
\label{sec:conclusions}
In this paper, we introduced a new data-informed approach for Reynolds stress closure.  Our approach is based on learning the components of the discrepancy tensor associated with a given Reynolds stress model with respect to the mean strain-rate eigenframe.  We first demonstrated through example that these components are smooth and hence relatively simple to learn using state-of-the-art supervised machine learning strategies and regression techniques.  We then demonstrated how to construct Galilean invariant, frame invariant, scale invariant, and computable discrepancy models by representing the components of the discrepancy tensor in the mean strain-rate eigenframe using artificial neural networks (ANNs) and learning the weights and biases of these ANNs using high-fidelity DNS data.  We refer to models built in this manner as ANN $S$-frame discrepancy models.  We continued by examining the energy stability properties of ANN $S$-frame discrepancy models, and we presented procedures for arriving at ANN $S$-frame discrepancy models which result in a global decay of mean kinetic energy and/or are realizable.  We next evaluated the effectiveness of ANN $S$-frame discrepancy models using four turbulent benchmark problems, turbulent channel flow, turbulent flow over a two-dimensional bump, a turbulent separation bubble created by suction and blowing, and turbulent flow over a backward facing step.  In particular, we constructed a universal $S$-frame discrepancy model using DNS data from all four turbulent benchmark problems that was substantially more accurate than the Spalart-Allmaras turbulence model.  Finally, we discussed how one could perform dimensional reduction of the input space for an ANN $S$-frame discrepancy model using Sobol’s total sensitivity indices.

There are several directions that we propose to explore in future work.  First, we will analyze the ability of ANN $S$-frame discrepancy models trained on a particular set of turbulent flows to extrapolate to other flows of interest with similar flow features.  The recent DNS database of flows over periodic hills obtained by the research group of Xiao could be employed for this purpose \cite{xiao2020flows}.  Second, we will apply ANN $S$-frame discrepancy models to more challenging turbulent flow applications, in particular, applications with unsteady and three-dimensional mean velocity and pressure fields.  Third, we will examine the impact of including other model inputs, such as distance to the wall or wall curvature, on the accuracy of ANN $S$-frame discrepancy models.  Fourth, we will pursue the construction of explicit analytical $S$-frame discrepancy models using the sensitivities of ANN $S$-frame discrepancy model with respect to model inputs.  Finally, we will pursue the development of data-driven approaches for learning the discrepancies between the model and exact closure terms appearing in model transport equations associated with a given Reynolds stress model, and we will combine said approaches with ANN $S$-frame discrepancy models to arrive at fully closed data-driven RANS models.  It should be noted that a fully closed data-driven RANS model may also be attained by constructing an ANN $S$-frame discrepancy model with inputs obtained from a state-of-the-art but imperfect RANS model, as in \cite{wang2017physics,wu2018physics}.  However, there are a number of potential issues associated with such a model.  For instance, it is anticipated that such a model will not always be predictive, especially for turbulent flow applications with both attached and separating regions.  Moreover, such a model cannot be used in the design of explicit analytical RANS models since the sensitivities of the data-driven model are with respect to imperfect RANS model inputs rather than the desired mean flow field and turbulence variables, and such a model is also unsuitable for use in hybrid RANS/LES simulations \cite{frohlich2008hybrid}.  We believe that the incorporation of data-driven RANS closures in hybrid RANS/LES simulations is a particularly attractive direction, as there likely exists no universally accurate Reynolds stress closure that depends only locally on mean flow quantities and turbulence variables \cite{spalart2015philosophies}.

\section{Acknowledgements}
This material is based upon work supported by the National Science Foundation under Grant CBET-1710670.  The authors would also like to thank Philippe Spalart and Rebecca Morrison, whose discussions improved upon the quality of this paper, and Basu Parmar and Aviral Prakash, who carefully proofread the paper.

\bibliographystyle{unsrt}
\bibliography{main}

\end{document}